\documentclass[10pt,twoside,a4paper]{article}
\usepackage{epigraph}
\usepackage{setspace}
\usepackage[pdftex]{graphicx}

\usepackage[scaled=.90]{helvet}
\usepackage[margin=1in, paperwidth=16.9cm, paperheight=23.9cm]{geometry} 
\usepackage[applemac]{inputenc} 

\usepackage{abstract}

\newcounter{qcounter}
\usepackage{filecontents}
\usepackage[defernums=true,isbn=false,maxnames=6,minnames=5,sorting=none]{biblatex}
\usepackage[nottoc,numbib]{tocbibind} 
\usepackage{placeins}

\usepackage{fourier-orns} 

\renewbibmacro{in:}{}

\usepackage{fancyhdr}
\usepackage{truncate}

\usepackage[flushleft]{threeparttable} 

\usepackage{nomencl}
\makenomenclature

\bibliography{thesis_refs}

\begin{document}

\begin{titlepage}

\setlength{\oddsidemargin}{0.5\paperwidth}
\addtolength{\oddsidemargin}{-0.5\textwidth}
\addtolength{\oddsidemargin}{-1in}
\setlength{\evensidemargin}{\oddsidemargin}
    
\begin{center}

\begin{spacing}{1}
\large Department of Cell and Molecular Biology\\
Karolinska Institutet, Stockholm, Sweden\\[4.0cm]
\end{spacing}

\begin{spacing}{2.0}
{ \sffamily\huge \bfseries Genome and transcriptome studies of the protozoan parasites \emph{Trypanosoma cruzi} and \emph{Giardia intestinalis}}\\[2.5cm]
\end{spacing}

\large Oscar Franz\'{e}n\\[2.5cm]

\includegraphics[width=0.6\textwidth]{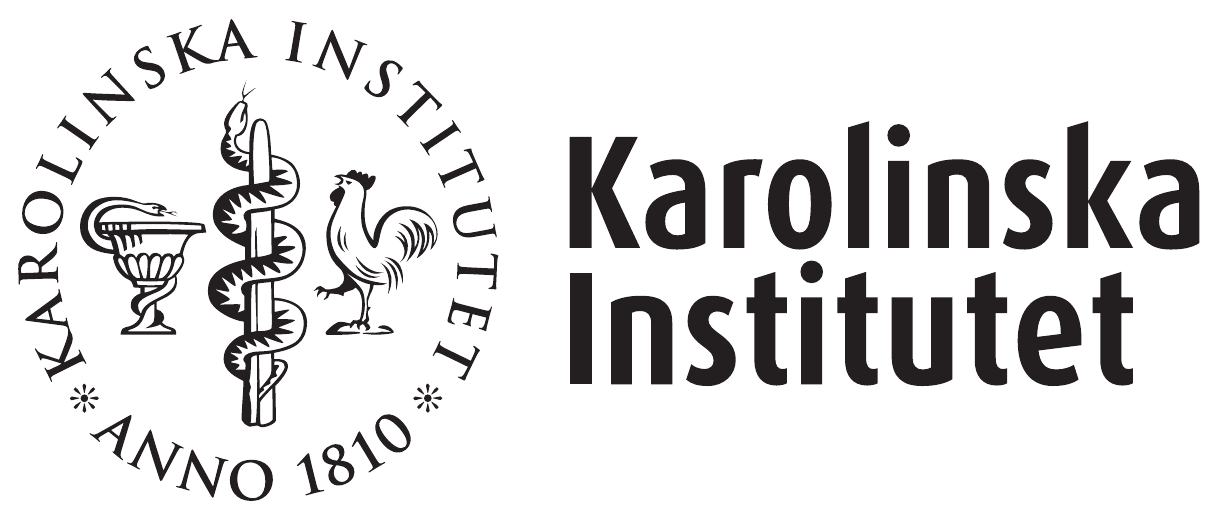}\\[1.5cm]

Stockholm 2012

\end{center}
\end{titlepage}

\thispagestyle{empty} 
\null 
\vfill
\noindent
\emph{Cover}: (Front) Artwork showing replicating \emph{Giardia intestinalis} trophozoites and \emph{Trypanosoma cruzi} epimastigotes. DAPI staining (blue) shows the two nuclei of the former parasite. Original micrographs: J. Jerlström-Hultqvist and M. Ferella.
\newline
\newline
\noindent
All previously published papers were reproduced with permission from the publisher.
\newline

\noindent
Published by Karolinska Institutet. Printed by Larserics Digital Print AB.
\newline

\noindent
© Oscar Franzén, 2012
\newline
\noindent

ISBN 978-91-7457-860-7
\pagebreak
\null

\null
\hfill 
\thispagestyle{empty} 
\emph{To my parents...}
\pagebreak 

\newpage
\mbox{}
\thispagestyle{empty} 

\clearpage
\thispagestyle{empty} 

\setlength\absleftindent{0in}
\setlength\absrightindent{0.3in}

\begin{onecolabstract}
\vspace*{-3.5cm}

\begin{spacing}{1.05}
\emph{Trypanosoma cruzi} and \emph{Giardia intestinalis} are two human pathogens and protozoan parasites responsible for the diseases Chagas disease and giardiasis, respectively. Both diseases cause suffering and illness in several million individuals. The former disease occurs primarily in South America and Central America, and the latter disease occurs worldwide.  Current therapeutics are toxic and lack efficacy, and potential vaccines are far from the market. Increased knowledge about the biology of these parasites is essential for drug and vaccine development, and new diagnostic tests. In this thesis, high-throughput sequencing was applied together with extensive bioinformatic analyses to yield insights into the biology and evolution of \emph{Trypanosoma cruzi} and \emph{Giardia intestinalis}. Bioinformatics analysis of DNA and RNA sequences was performed to identify features that may be of importance for parasite biology and functional characterization.
This thesis is based on five papers (\emph{i}-\emph{v}). Paper \emph{i} and \emph{ii} describe comparative genome studies of three distinct genotypes of \emph{Giardia intestinalis} (A, B and E). The genome-wide divergence between A and B was 23\% and 13\% between A and E. 4557 groups of three-way orthologs were defined across the three genomes. 5 to 38 genotype-specific genes were identified, along with genomic rearrangements. Genes encoding surface antigens, \emph{vsps}, had undergone extensive diversification in the three genotypes. Several bacterial gene transfers were identified, one of which encoded an acetyltransferase protein in the E genotype. Paper \emph{iii} describes a genome comparison of the human infecting \emph{Trypanosoma cruzi} with the bat-restricted subspecies \emph{Trypanosoma cruzi marinkellei}. The human infecting parasite had an 11\% larger genome, and was found to have expanded repertoires of sequences related to surface antigens. The two parasites had a shared `core' gene complement. One recent horizontal gene transfer was identified in \emph{T. c. marinkellei}, representing a eukaryote-to-eukaryote transfer from a photosynthesizing organism. Paper \emph{iv} describes the repertoire of small non-coding RNAs in \emph{Trypanosoma cruzi} epimastigotes. Sequenced small RNAs were in the size range 16 to 61 nucleotides, and the majority were derived from transfer RNAs and other non-coding RNAs. 92 novel transcribed loci were identified in the genome, 79 of which were without similarity to known RNA classes. One population of small RNAs were derived from protein-coding genes. Paper \emph{v} describes transcriptome analysis using paired-end RNA-Seq of three distinct genotypes of \emph{Giardia intestinalis} (A, B and E). Gene expression profiles recapitulated the known phylogeny of the examined genotypes, and 61 to 176 genes were differentially expressed. 49,027 distinct polyadenylation sites were mapped and compared, and the median 3$'$UTR length was 80 nucleotides (A). One 36-nt novel intron was identified and the previously reported introns (5) were confirmed.
\end{spacing}
\end{onecolabstract}
\pagebreak

\thispagestyle{empty} 
\newpage{} 

\section*{List of Publications}

This thesis is based on the following articles and manuscripts that will be referred to in the text by their roman numerals.

\begin{list}
{\roman{qcounter}~}
{
	\usecounter{qcounter}
	\setlength\labelwidth{3in}
	\setlength\listparindent{0.5in}
	\setlength\parsep{0in}
}

\item \textbf{Franzén O}, Jerlström-Hultqvist J, Castro E, Sherwood E, Ankarklev J, Reiner DS, Palm D, Andersson JO, Andersson B, Svärd SG. \textbf{Draft Genome Sequencing of \emph{Giardia intestinalis} Assemblage B Isolate GS: Is Human Giardiasis Caused by Two Different Species?}. \emph{PLoS Pathogens, 2009, Aug;5(8):e1000560}.
\item Jerlström-Hultqvist J, \textbf{Franzén O}, Ankarklev J, Xu F, Noh\'{y}nková E, Andersson JO, Svärd SG, Andersson B. \textbf{Genome analysis and comparative genomics of a \emph{Giardia intestinalis} assemblage E isolate}. \emph{BMC Genomics, 2010, 11:543}.
\item \textbf{Franzén O}, Talavera-López C, Ochaya S, Butler CE, Messenger LA, Lewis MD, Llewellyn MS, Marinkelle CJ, Tyler KM, Miles MA, Andersson B. \textbf{Comparative genomic analysis of human infective \emph{Trypanosoma cruzi} lineages with the bat-restricted subspecies \emph{T. cruzi marinkellei}}. \emph{BMC Genomics}. Accepted (in press).
\item \textbf{Franzén O} $^{\dagger}$, Arner E $^{\dagger}$, Ferella M, Nilsson D, Respuela P, Carninci P, Hayashizaki Y, Åslund L, Andersson B, Daub CO.  \textbf{The Short Non-Coding Transcriptome of the Protozoan Parasite \emph{Trypanosoma cruzi}}. \emph{PLoS Neglected Tropical Diseases, 2011, 5(8): e1283}.\newline
\small $^{\dagger}$ Equal contribution.
\normalsize \item \textbf{Franzén O} $^{\ddagger}$, Jerlström-Hultqvist J $^{\ddagger}$, Einarsson E, Ankarklev J, Ferella M, Andersson B, Svärd SG. \textbf{Transcriptome Profiling of \emph{Giardia intestinalis} Using Strand-specific RNA-Seq}. \emph{Submitted manuscript}.\newline
\small $^{\ddagger}$ Equal contribution.
\end{list}

\thispagestyle{empty} 
\newpage{} 
\section*{Other Publications}

\thispagestyle{empty} 

During the course of my doctoral studies, I also performed or participated in the following studies.

\begin{itemize}
	\item \textbf{Franzén O}, Ochaya O, Sherwood E, Lewis MD, Llewellyn MS, Miles MA, Andersson B. \textbf{Shotgun Sequencing Analysis of \emph{Trypanosoma cruzi} I Sylvio X10/1 and Comparison with \emph{T. cruzi} VI CL Brener}. \emph{PLoS Neglected Tropical Diseases, 2010, 5(3): e984}.
	\item Jiao W, Masich S, \textbf{Franzén O}, Shupliakov O. \textbf{Two pools of vesicles associated with the presynaptic cytosolic projection in \emph{Drosophila} neuromuscular junctions}. \emph{Journal of Structural Biology, 2010, 389–394}.
	\item Messenger LA, Llewellyn MS, Bhattacharyya T, \textbf{Franzén O}, Lewis MD, Ramírez JD, Carrasco HJ, Andersson B, Miles MA. \textbf{Multiple Mitochondrial Introgression Events and Heteroplasmy in \emph{Trypanosoma cruzi} Revealed by Maxicircle MLST and Next Generation Sequencing}. \emph{PLoS Neglected Tropical Diseases, 2012, 6(4): e1584}.
\end{itemize}

\newpage
\mbox{}
\thispagestyle{empty} 
\newpage

\clearpage
\thispagestyle{empty} 
\tableofcontents
\thispagestyle{empty}
\newpage{} 

\nomenclature{CDS}{Coding sequence}
\nomenclature{ORF}{Open reading frame}
\nomenclature{DNA}{Deoxyribonucleic acid}
\nomenclature{bp}{Base pairs}
\nomenclature{SGS}{Second generation sequencing}
\nomenclature{contig}{Contiguous sequence}
\nomenclature{FPKM}{Fragments per kilobase of transcript per million fragments mapped}
\nomenclature{PAC}{Polyadenylation site cluster}
\nomenclature{SAGE}{Serial analysis of gene expression}
\nomenclature{RNA-Seq}{RNA-Sequencing}
\nomenclature{ASE}{Allele-specific expression}
\nomenclature{AER}{Allelic expression ratio}
\nomenclature{AR}{Allele ratio}
\nomenclature{cDNA}{Complementary DNA}
\nomenclature{LBA}{Long branch attraction}
\nomenclature{PCR}{Polymerase chain reaction}
\nomenclature{DTU}{Discrete typing unit}
\nomenclature{PTU}{Polycistronic transcription unit}
\nomenclature{RNAi}{RNA interference}
\nomenclature{SNP}{Single nucleotide polymorphism}
\nomenclature{sRNA}{Small RNA}
\nomenclature{RNA}{Ribonucleic acid}
\nomenclature{nt}{Nucleotides}
\nomenclature{tsRNA}{tRNA-derived small RNA}
\nomenclature{UTR}{Untranslated region}
\nomenclature{polyA}{Polyadenylation}
\nomenclature{s.s.}{sensu stricto}

\setlength{\nomitemsep}{-\parsep} 

\printnomenclature
\thispagestyle{empty} 
\newpage{} 


\newpage
\mbox{}
\thispagestyle{empty} 
\newpage


\setcounter{page}{1}

\section{Synopsis}

\fancyhead[RO,LE]{}

\setlength\epigraphwidth{3.6in}
\setlength{\epigraphrule}{0pt} 

\epigraph{\emph{Nothing in biology makes sense except in the light of evolution\newline}}{-- T.G. Dobzhansky (1973), geneticist and evolutionary biologist.}

\vspace{1.0cm}

Infectious diseases are leading causes of suffering and death of humans around the world, and have significant impact on daily lives of many million people. Parasites cause some of the worst and most neglected diseases, including Malaria, African- and American trypanosomiasis, Schistosomiasis and several others. These diseases are most prevalent in tropical regions of the world, and are often associated with poverty as well as being intrinsically ``poverty promoting.'' Unsafe drinking water, compromised hygiene and sanitary conditions or substandard housing are factors that facilitate disease. Many of the afflicted individuals have very limited access to health care. Several factors can be attributed to the lack of treatment options: neglected diseases attract little attention from pharmaceutical companies and first-world governments, often because companies are unable to regain investments in expensive basic research, drug development and clinical trials. Moreover, many parasites are difficult to study in the laboratory due to complex life cycles or because the parasites do not readily grow \emph{in vitro}. Most neglected diseases do not cause acute outbreaks, and instead progress during many years and in the meantime cause debilitating illness and suffering.

In addition to the mission of improving human health, parasites often have unique or specialized biological features, which makes them excellent models for the study of eukaryotic evolution. Parasites provide a window into the biological and social evolution of our own species, since many parasites have co-evolved together with the \emph{Homo} lineage for many millions of years; this appears to be the situation for the worms \emph{Trichuris trichiura} and \emph{Enterobius} \cite{Anderson1997,Le2000}; other parasites such as \emph{Trypanosoma cruzi} tell a shorter story of human co-evolution. Parasite-host co-evolution has likely resulted in reciprocal adaptations with complex evolutionary consequences, for example the favouring of specific genetic processes such as recombination that operate to create new genotypes to which the host is not adapted. Another example of co-evolution can be found in our own species, where a trypanolytic factor encoded by our genome is likely an ancestral adaptation against trypanosomatid parasites of the \emph{T. brucei} clade \cite{Pays2006}.

Second generation sequencing enables cost-efficient and rapid acquisition of large data sets covering  diverse biological aspects, including but not limited to genome, metagenome, epigenome and transcriptome  studies. An important target of the new technologies is human parasites, aiming to deepen our understanding of the underlying biology of these pathogens, and their evolutionary trajectories. Such efforts may reveal signatures relating to how parasitism evolved. Second generation sequencing has already facilitated key insights into the molecular organization of these organisms, and rapidly enables advancement of functional studies, identification of drug targets and formation of new hypotheses. Old questions relating to pathogenicity, epidemiology and genetics can be addressed with the new tools and may ultimately lead to insights that pave the way for better treatment strategies for neglected and tropical diseases.

\begin{center}
\decofourleft\decofourright
\end{center}

\newpage 


\null
\thispagestyle{empty} 

\topskip0pt

\section{Introduction}

\fancyhead[RO,LE]{Genome and RNA sequencing}

\vspace{0.1cm}

\setlength\epigraphwidth{3.5in}
\setlength{\epigraphrule}{0pt} 

\epigraph{\emph{Fossil bones and footsteps and ruined homes are the solid facts of history, but the surest hints, the most enduring signs, lie in those miniscule genes. For a moment we protect them with our lives, then like relay runners with a baton, we pass them on to be carried by our descendents. There is a poetry in genetics which is more difficult to discern in broken bones, and genes are the only unbroken living thread that weaves back and forth through all those boneyards.\newline}}{-- J. Kingdon, biologist and science author (1996).}

\subsection{Current State of Genome Sequencing}
\begin{figure}[!ht]
 \centering
 \includegraphics[width=4.5in]{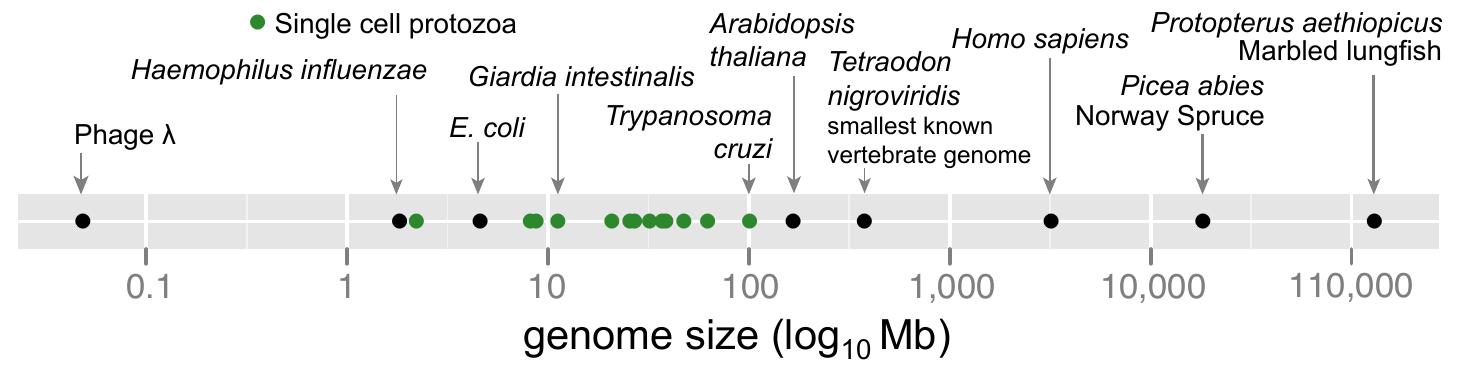}
  \caption{ \small Genome sizes on a logarithmic scale. Phage $\lambda$ has a genome of 48,502 bp and it was determined by Sanger \emph{et al.} in 1982 \cite{Sanger1982}. Most genomes of sequenced, parasitic protozoa are between 4 to 100 Mb in size. The genome size of \emph{T. cruzi} refers to both haplotypes of the CL Brener strain. For the other eukaryotes the genome size refers to the haploid state. (Green dots) Single cell protozoa. Genomes of mammals are at least an order of magnitude larger. Marbled lungfish has the largest genome of any known organism (130,000 Mb) \cite{Metcalfe2012}. }
  
\label{genome_size_plot}
\end{figure}

\begin{figure}[!ht]
 \centering
 \includegraphics[width=4.5in]{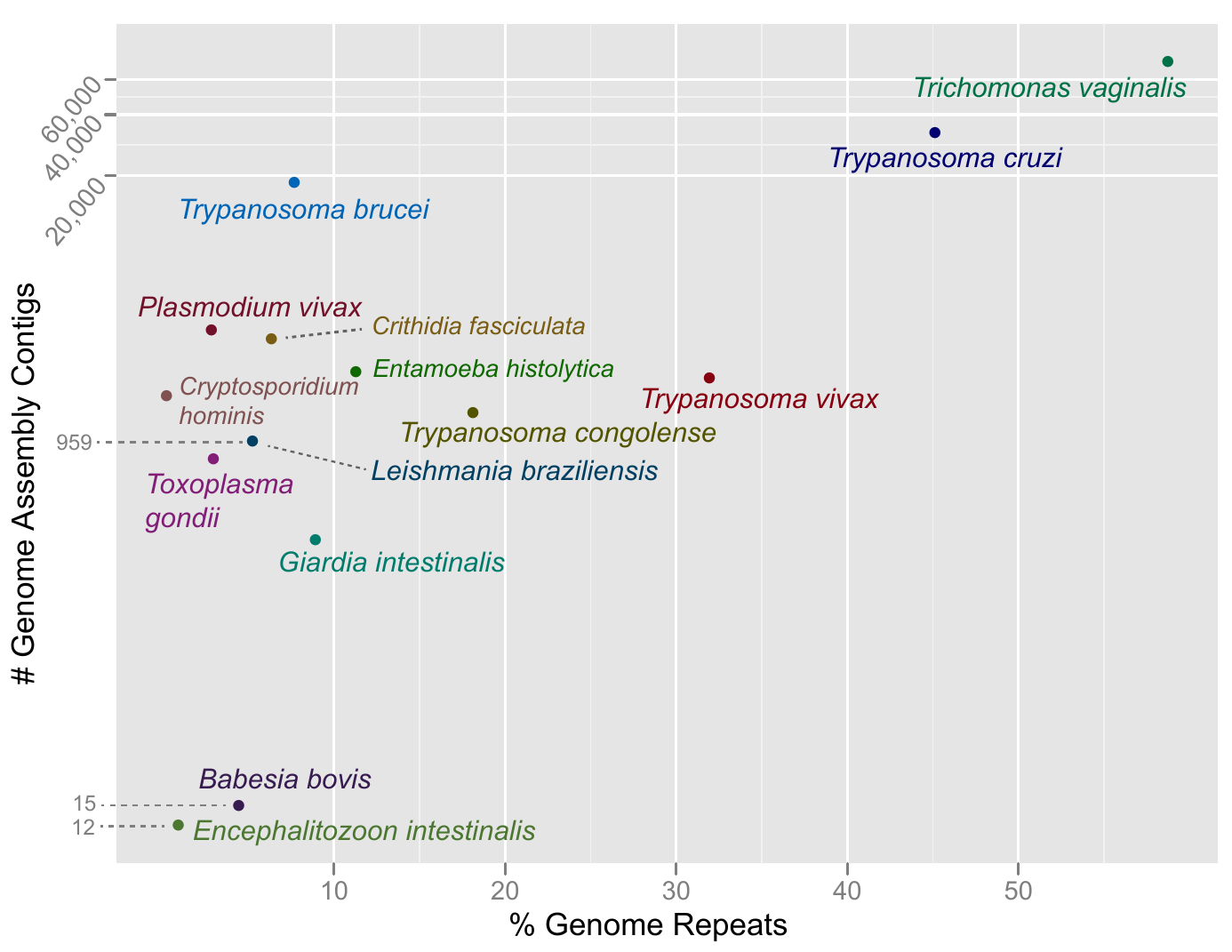}
  \caption{ \small Comparison of genome repeat-content and assembly fragmentation of various eukaryotic pathogens from EupathDB \cite{Aurrecoechea2010}. Repeat libraries were established for each genome using RepeatScout \cite{Price2005} (repeats \textgreater 500 bp of length), and contigs \textgreater 200 bp of each genome were searched using RepeatMasker \cite{TarailoGraovac2009}. (X-axis) Percentage of the genome (sum of contig lengths) present in repeats. (Y-axis) Contig count of the assembly (logarithmic scale). Strains: \emph{B. bovis} (T2Bo); \emph{C. fasciculata} (Cf-C1); \emph{C. hominis} (TU502); \emph{E. intestinalis} (ATCC 50506); \emph{E. histolytica} (HM-1:IMSS); \emph{G. intestinalis} (WB); \emph{L. braziliensis} (M2903); \emph{P. vivax} (SaI-1); \emph{T. gondii} (GT1); \emph{T. vaginalis} (G3); \emph{T. brucei} (427); \emph{T. congolense} (IL3000); \emph{T. cruzi} (CL Brener); and \emph{T. vivax} (Y486). }

\label{genome_repeats}
\end{figure}

Genomes vary enormously in size, with many of the size differences being caused by repeated sequences (Figure \ref{genome_size_plot}). Deoxyribonucleic acid (DNA) sequencing techniques are limited to reading short pieces of DNA. The most common strategy to overcome this limitation is shotgun sequencing. The technique involves random fragmentation of chromosomes to a redundant mix of small DNA fragments. These fragments can subsequently be sequenced, resulting in a mix of sequences of forward and reverse directions, representing the original chromosomes. By using the overlap of these short sequences, computer programs can reconstruct millions of short sequences into longer sequences (contigs) -- a process referred to as genome assembly. While simple in theory, the task becomes less straightforward due to the following obstacles: (\emph{i}) the large size of most genomes, especially those of eukaryotes; (\emph{ii}) the vast amount of sequence data needed to achieve sufficient redundancy, i.e. the genome ``coverage''; (\emph{iii}) the fact that many DNA fragments are identical or close to identical (repeats); (\emph{iv})  heterozygosity, i.e. single nucleotide polymorphisms between homologous chromosomes; and (\emph{v}) sequence errors. Apart from these issues, genomes can exhibit aneuploidy and complex karyotypes -- all of which make the assembly task more difficult. Sanger sequencing is referred to as `first generation sequencing,' and has been used to sequence large and complex genomes, e.g. the human genome. Sanger sequencing can read up to 1000 base pairs (bp) of a DNA fragment. Despite being relatively old, it is still the most common technique for low-throughput applications, e.g. DNA amplified from Polymerase Chain Reaction (PCR). A comparison of the repeat content in relation to assembly consistency of various draft or complete genomes is shown in Figure \ref{genome_repeats}.

\FloatBarrier

Second generation sequencing (SGS) offers higher throughput, at lower cost per base, but yields shorter sequences (reads). Short reads are often a problem for determining a genome sequence, as most genomes contain repetitive sequences longer than the read length \cite{Alkan2011}. Paired-end protocols have been developed to tackle repeats, and allow sequencing of longer DNA fragments from both ends in order to bridge repeats. Paired-end reads of various sizes can subsequently be used to link contigs into scaffolds. Hence, `scaffolds' is the genomics term for contigs that are ordered and oriented. Several SGS techniques or platforms are available, for example Roche/454 sequencing, developed from pyrosequencing. The platform from Illumina provides significantly higher throughput, albeit at shorter read lengths. Most genome sequencing efforts combine data from different platforms to overcome their respective limitations. Repetitive sequences are currently the major bottleneck in genome sequencing projects, especially since most eukaryotes contain various classes of repeats, e.g. retroelements and segmental duplications. Future developments are anticipated to improve genome sequences, including Pacific Biosciences and perhaps more distant, Nanopore sequencing \cite{Wanunu}.

\subsection{RNA-Seq -- A Method to Read the Transcriptome at Single Nucleotide Resolution}
The field of transcriptomics aims to describe all transcripts in a cell or tissue and to determine the features of these transcripts: for example 5$'$ and 3$'$ end structures, splicing patterns and quantitative information about transcript levels. Transcriptome sequencing (RNA-Seq) relies on deep sequencing of fragmented cDNA libraries, achieving its quantitative properties by the amounts of sequence reads of a particular transcript \cite{Wang2009}, i.e. abundant transcripts yield more reads whereas more rare transcripts yield fewer. Therefore, the final digital expression values are based on simple counting statistics, which are often normalized by gene length and the number of sequences generated by the instrument. While microarray data often require complicated normalization procedures, processing of RNA-Seq data is relatively simple and straightforward. RNA-Seq results in lower background noise than microarrays \cite{Fu2009}. Any high-throughput sequencing platform can be utilized, but Illumina has been the most widely adopted and is currently the best supported in terms of bioinformatics software. RNA-Seq enables \emph{ab initio} discovery of new and rare transcripts and splicing patterns, which cannot be observed on standard microarrays. Several software programs are freely available to process RNA-Seq data. These programs rely on optimized algorithms to map (align) large quantities of sequence data, and at the same time consider polymorphisms and sequence errors. RNA-Seq has rapidly become widely adopted and applied to various research questions, including differential expression analysis \cite{Robinson2010}, small ribonucleic acid (RNA) discovery, allele-specific expression \cite{Degner2009} as well as mapping 5$'$ and 3$'$ ends of genes \cite{Shepard2011}.

Recent developments allow the generation of paired-end libraries, read lengths up to 150 nucleotides and strand specificity. These improvements allow going beyond the mRNA component of the transcriptome and sampling hidden transcriptional layers. Drawbacks of the RNA-Seq method include: (\emph{i}) the cost of library preparation and sequencing; (\emph{ii}) lack of user-friendly analysis pipelines and interfaces; (\emph{iii}) RNA or cDNA must be fragmented into smaller pieces, usually between 100 to 500 nt; (\emph{iv}) library preparation and  fragment amplification may introduce artifacts or biases; (\emph{v}) transcript coverage bias is common, i.e. coverage fluctuations along the 5$'$ to 3$'$ axis of the mRNA; (\emph{vi}) long-time storage of RNA-Seq data sets is becoming increasingly difficult because of large data volumes; and (\emph{vii}) certain downstream analysis tasks, e.g. discovery of rare transcript isoforms and splice variants, still suffer from many false positives due to artifactual chimeras or amplification biases from library preparation or sequencing. Future developments in single molecule sequencing may ameliorate such problems. Despite these limitations, RNA-Seq is likely to improve and provide novel insight into the transcriptomes of protozoans and other eukaryotes.

\begin{center}
\decofourleft\decofourright
\end{center}

\newpage

\subsection{\emph{Giardia intestinalis} -- A Gastrointestinal Parasite of Humans and Animals}
\fancyhead[RO,LE]{\emph{Giardia intestinalis}}

\emph{Giardia intestinalis} is a protozoan, amitochondrial parasite and member of the diplomonad group of species, which includes other anaerobic or microaerophilic protozoans. The diplomonad group is part of the supergroup Excavata \cite{Simpson2003}. In the literature, the species names \emph{G. intestinalis}, \emph{G. duodenalis} and \emph{G. lamblia} are used interchangeably and refer to the same organism. The parasite was discovered already in 1681 by the Dutch microscopist van Leeuwenhoek \cite{Dobell1920}, and described in more detail in 1859 by the Czech physician Lambl \cite{Lambl1859}. \emph{G. intestinalis} infects humans and animals and is one of the most prevalent gastrointestinal parasites worldwide \cite{Adam2001}. \emph{G. intestinalis} is a potential zoonotic pathogen since it can infect a broad range of mammals in addition to humans. In man, the parasite colonizes the upper part of the small intestine and adheres to the mucosa along the sides of villi. The infection causes diarrhea, and may lead to malnutrition and failure of children to thrive \cite{Ankarklev2010}. The disease is particularly a burden in developing countries, where compromised hygiene may increase transmission and cause endemic outbreaks. Local outbreaks do occasionally occur in developed countries, for example via the public water supply \cite{Kent1988} or in day care centers \cite{Black1977}. In 2004 a large outbreak of \emph{G. intestinalis} occurred in Bergen Norway, with altogether 1,300 laboratory-confirmed cases \cite{Nygard2006}. Since the outbreak certain individuals have had prolonged and recurring symptoms of giardiasis, with a profound impact on the quality of life \cite{Robertson2010}. Recent data indicate a putative relationship between irritable bowel syndrome and previous \emph{G. intestinalis} infection \cite{Wensaas2012}.

As of 2004, \emph{G. intestinalis} has been included in the WHO Neglected Disease Initiative \cite{Savioli2006}.

\subsubsection{Cell Biology and Life Cycle: Regression and Simplicity}
\begin{figure}[!ht]
 \centering
 \includegraphics[width=3in]{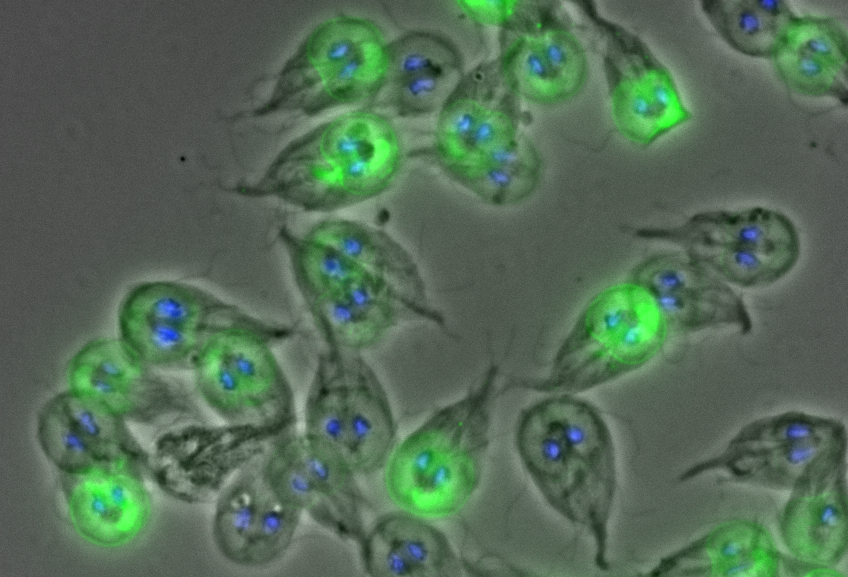}
  \caption{ \small \emph{G. intestinalis} trophozoites seen through a microscope. (Green) Tagged median body protein. (Blue) DAPI stained nuclei. Image credit: J. Jerlström-Hultqvist. }
  
\label{trops}
\end{figure}

In contrast to other protozoan parasites, \emph{G. intestinalis} has a relatively simple life cycle, consisting of the dormant cyst stage and the replicative trophozoite stage. Trophozoites have a characteristic half pear-shaped morphology, and are 12-15 $\mu$m long, 5-9 $\mu$m wide and 1-2 $\mu$m thick (Figure \ref{trops}). Trophozoites have four pairs of flagella, which are anchored to the cytoskeleton. The parasite rotates around its longitudinal axis to create a forward propulsion force. The rotation causes the parasite to move at a speed of 12-40 $\mu$m/s \cite{Lenaghan2011}. An adhesion disk is present on the ventral surface of the parasite, and is used for anchoring to substrates. Hernández-Sánchez \emph{et al.} reported that adhesion-deficient \emph{G. intestinalis} had reduced capacity to establish infection in Mongolian gerbils \cite{HernandezSanchez2008}. When the parasite anchors to substrates, whether artificial or natural, the pattern of motion changes to more stable planar swimming \cite{Lenaghan2011}. Unusually compared to most eukaryotes, trophozoite cells contain two transcriptionally active nuclei \cite{Ankarklev2010}. Each nucleus contains a diploid to tetraploid set of the genome \cite{Bernander2001}. The biological significance of the polyploid genome is not clear, but it is a shared feature among many diplomonads (order \emph{Diplomonadida}) and likely relates to the evolutionary history of the order. \emph{G. intestinalis} has a well-defined endoplasmatic reticulum, which can form excretory vesicles \cite{Abodeely2009}. \emph{G. intestinalis} lacks a canonical Golgi apparatus and mitochondria. A vestigial organelle called mitosome is present. Mitosomes are double-membraned structures that appear to be involved in iron metabolism \cite{Tovar2003}. 

\FloatBarrier

Cysts are the non-motile and metabolically dormant stage of the life cycle, and represent the infectious agents of giardiasis. Cysts can persist in the environment for prolonged periods and remain infectious, being encapsulated in a thick cyst wall of carbohydrate and protein. The most common route of transmission is the fecal-oral route, via contaminated food or water. Infection can also occur via person-to-person contact due to poor hygiene. The infectious dose can be as low as 10 cysts, as shown in a Texas prison ``volunteer'' population in 1954 \cite{Rendtorff1954}. Ingested cysts undergo excystation inside the host, triggered by stomach acids. Cysts then rupture in the small intestine. Giardiasis is characterized by watery diarrhea, gastric pain and weight loss, and often but not always resolves spontaneously. The pathophysiology of giardiasis is poorly understood, but likely involves dysfunction of the epithelial cell barrier of the intestine and disturbances of the electrolyte balance \cite{Troeger2007}. The infection triggers apoptosis of host epithelial cells, and causes shortening of the brush border villi, all of which may contribute to diarrhoea \cite{Ankarklev2010}. There is also evidence that proteolytic enzymes released by \emph{G. intestinalis} are involved in the disease \cite{deCarvalho2008}. Encystation is the process where trophozoites are transformed back to cysts, and it is triggered by the intestinal environment (high levels of bile, low cholesterol and/or shift in pH) \cite{Ankarklev2010}.

\subsubsection{Is \emph{G. intestinalis} a Primitive Eukaryote or Highly Adapted Towards Parasitism?}
Phylogenies based on nucleotide and protein sequences have consistently identified \emph{G. intestinalis} as a basal eukaryote \cite{Sogin1989,Leipe1993,Hashimoto1995,Hilario1998}. This view has been corroborated by the apparent lack of some intracellular compartments (e.g. mitochondria, Golgi and peroxisomes) and an overall simplified, bacterial-like metabolism \cite{Adam2001}.

Roger \emph{et al.} reported the finding of the mitochondrial-like gene \emph{cpn60} in the genome of \emph{G. intestinalis} \cite{Roger1998}. The same year Hashimoto \emph{et al.} reported the finding of a nuclear-encoded valyl-tRNA synthetase gene \cite{Hashimoto1998}, which is regarded to be of mitochondrial origin in eukaryotes. A mitochondria-derived organelle, the mitosome, was later discovered \cite{Tovar2003}. Together these data suggest that \emph{G. intestinalis} diverged after the endosymbiosis of the mitochondrial ancestor, but subsequently lost this feature, possibly as an adaptation to the microaerophilic life in the intestine. The finding of nucleoli also points in the direction of a typical eukaryote \cite{JimenezGarcia2008}. The early-branching position of \emph{G. intestinalis} in phylogenetic trees has been questioned as an artifact caused by long-branch attraction (LBA) \cite{Dacks2002}. The problem of LBA arises when comparing taxa with variable evolutionary rates, which may lead to the artifactual early emergence of these taxa \cite{Brinkmann2005}. The effect of LBA may be mitigated by inclusion of additional species. Analysis of small nucleolar RNAs from Archaea and various unicellular eukaryotes has suggested that \emph{G. intestinalis} emerged later than \emph{Trypanosoma} and \emph{Euglena} \cite{Luo2009}. Altogether the current data of this parasite indicate that it has undergone reductive evolution and is highly adapted towards parasitism.

\subsubsection{Intraspecific Taxonomy: Two Genotypes Infect Humans}
\emph{G. intestinalis} propagates via binary fission, i.e. it is an asexual process. Whether \emph{G. intestinalis} participates in rare sexual events has been subject of debate, but there is currently no direct evidence for a sexual or parasexual cycle. Tibayrenc \emph{et al.} showed that the parasite meets the criteria for a predominantly clonal population structure \cite{Tibayrenc1990}. Nevertheless, recent data have suggested the possibility of infrequent genetic exchange \cite{Ramesh2005,LasekNesselquist2009}. Many human-infective protozoans have sexual cycles or infrequently participate in genetic exchange; including \emph{Trypanosoma cruzi} \cite{Gaunt2003}, \emph{Toxoplasma gondii} \cite{Black2000} and \emph{Plasmodium falciparum} \cite{Bruce1990}. Predominant clonal propagation does not preclude the existence of rare sexual events, but several questions are unresolved or inconsistent with a conventional sexual organism.

\begin{figure}[!ht]
 \centering
 \includegraphics[width=4.5in]{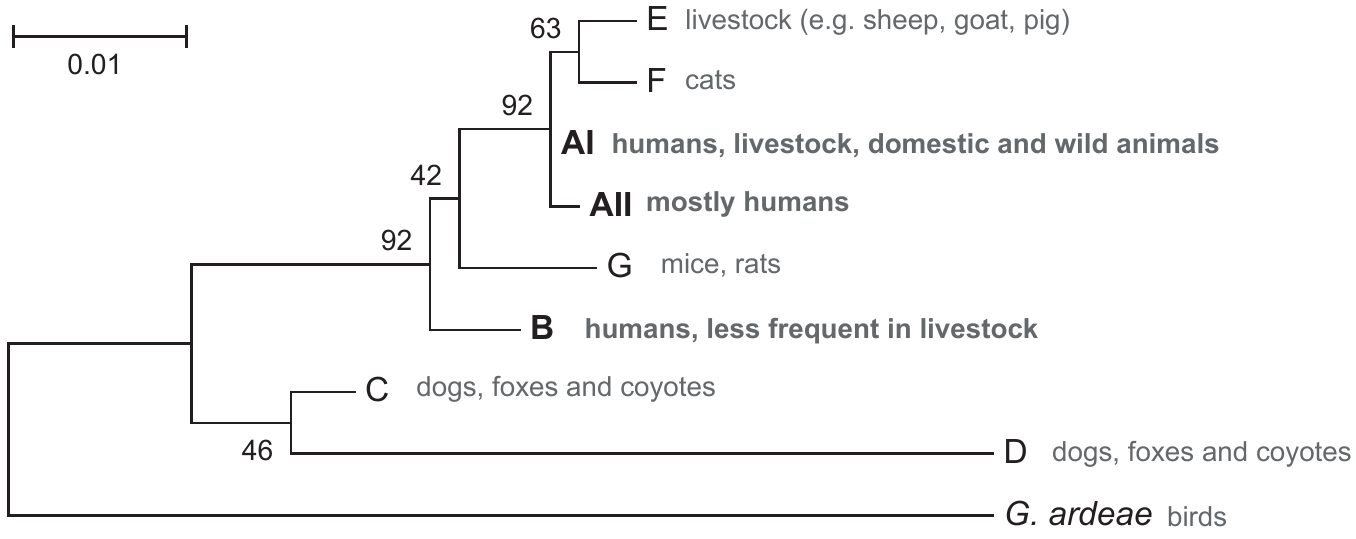}
  \caption{ \small Maximum likelihood phylogeny of \emph{G. intestinalis} assemblages A to G based on the \emph{ef1$\alpha$} gene (nucleotide sequences). Sequences were aligned with ClustalW2 and the topology was inferred using the Tamura-Nei model as implemented in MEGA5 \cite{Tamura2011}. The scale bar refers to substitutions per site. Support values at branches were generated from 1000 bootstrap replicates. \emph{G. ardeae}, a species found in birds \cite{Erlandsen1990}, was used as an outgroup to support the phylogeny. Accession numbers of  sequences used to infer the phylogeny; D14342.1, AF069573.1, AF069570.1, AF069574.1, AF069575.1, AF069571.1, AF069572.1, AF069568.1, AF069567.1. }
  
\label{giardia_phylogeny}
\end{figure}

\emph{G. intestinalis} has been suggested to comprise a species complex, consisting of eight distinct but morphologically indistinguishable genotypes (assemblages; Figure \ref{giardia_phylogeny}). Of the eight recognized assemblages (A to H), only two (A and B) infect humans as well as various non-human primates, cattle and many other animals \cite{Feng2011}. Population studies have revealed further substructure of assemblage A, which can be subgrouped into AI, AII and AIII. Variation in pathogenicity among strains has been documented \cite{Thompson2001,Sahagun2008}, indicating a putative relationship between genotype and symptomatology. However, attempts to associate genotype with disease outcome have often been conflicting, and there is currently no certain relationship. In contrast, assemblage B has no clear subgrouping \cite{Feng2011}. Only assemblage B has been used in experimental human infections \cite{Nash1987}.

Assemblages C to H are not associated with human infections, display stronger host-specificity and are less studied due to the difficulty to cultivate them \emph{in vitro}. Parasites from assemblages C and D have been identified in dogs, wolves, coyotes and cats; E parasites have been found in cattle, sheep, pigs, goats and water buffalo; F parasites are reported mainly in cats; G parasites are mainly in rodents \cite{Feng2011}. H parasites were relatively recently discovered in marine vertebrates \cite{LasekNesselquist2010}. The phylogenetic topology of the assemblages suggests that the extant A, E and F lineages share a common ancestor. It is possible that animal domestication provided opportunities for parasites to cross species boundaries and thereby adapt to new hosts.

\FloatBarrier

Because of the different host species as well as genetic characteristics, reorganization of the assemblages into separate species has been proposed and is under debate \cite{Monis2009}. The new species names for A and B are suggested to be \emph{G. duodenalis} and \emph{G. enterica}, respectively. Further studies and gathering of phenotypic data may shed light on the biology of non-human associated \emph{Giardia} parasites.

\subsubsection{The Streamlined Genome of \emph{G. intestinalis} Reveals Many Parasite-specific Genes}
One striking feature of \emph{G. intestinalis} is the highly reduced genome, which is comprised of {\raise.17ex\hbox{$\scriptstyle\sim$}}12 million base pairs (haploid size) distributed on five chromosomes \cite{Adam2001}. Upcroft \emph{et al.} reported a certain amount of karyotype variability in human and animal stocks \cite{Upcroft1989}, suggesting that the karyotype is not completely stable. Each nucleus of trophozoites contains a diploid to tetraploid set of the genome \cite{Kabnick1990,Yu_2002,Bernander2001}. The genome of the assemblage A isolate WB was finished 2007 \cite{Morrison2007}, and revealed sparse non-coding sequences, and with few exceptions intronless genes and simplified cellular components; including many bacterial- and archaeal-like enzymes. DNA synthesis, transcription, RNA processing and cell cycle components were found to be simple \cite{Morrison2007}. Since the completion of the genome sequence, five genes containing introns have been found \cite{Morrison2007,Nixon2002,Russell2005,Roy2012}. The discovery of splicing in \emph{G. intestinalis} was surprising and suggested that splicing was present at an early stage in the ancestral eukaryote. Recently, three \emph{trans}-splicing events have been uncovered, i.e. the joining of physically distant exons into contiguous mature mRNAs \cite{Kamikawa2011,Roy2012}. The documented events involved exons of the \emph{dhc} and \emph{hsp90} genes. The finding of such relatively complicated transcript maturation pathways further contradicts the view of \emph{G. intestinalis} as a ``fossil'' eukaryote. The genome does encode proteins involved in meiosis \cite{Morrison2007}, but these may have alternative functions.

Despite extensive efforts to annotate the \emph{G. intestinalis} genome, {\raise.17ex\hbox{$\scriptstyle\sim$}}58\% of the genes are without a known function. These genes lack sequence similarity to other sequenced genomes and may represent \emph{Giardia}-specific genes. The \emph{G. intestinalis} genome encodes four multigene families: \emph{nek} (encoding kinases), \emph{p21.1} (encoding structural proteins, containing ankyrin motifs), \emph{vsp} (surface antigens) and \emph{hcmp} (putative surface antigens). Altogether these genes comprise 30\% (3.6 Mbp/12 Mbp) of the genome \cite{Morrison2007}, indicating that gene duplication has been a major evolutionary force. Most genes of these families display extensive heterogeneity, indicating significant divergence since the presumed gene duplication event. 198 \emph{nek} genes are present in the genome, most of which are predicted to encode catalytically inert proteins \cite{Morrison2007}. Manning \emph{et al.} found some nek proteins localized to distinct parts of the cytoskeleton and cytoplasm \cite{Manning2011}. However, the precise roles of most of these proteins are unknown. Variant-specific surface antigens encoded by \emph{vsp} genes cover the surface of the parasite and shield it from the host immune system \cite{Adam2001}. Sequence analysis of \emph{vsp} genes has suggested substructure, recombination and divergence among these genes \cite{Adam2010}. Only one \emph{vsp} is expressed at the cell surface at any given time \cite{Nash2001}, and switching occurs every 6-13 generations \cite{Nash1990}. \emph{vsp} switching occurs spontaneously, and is proposed to involve an epigenetic mechanism \cite{Kulakova2006} and/or RNA interference \cite{Prucca2008}.

The \emph{G. intestinalis} genome contains three families of retrotransposons, of which two are potentially active and localized to telomeres and one is dead and present in interstitial genomic regions \cite{Arkhipova2001}. All three families of retrotransposons are long interspersed nuclear element (LINE)-like elements. Ullu \emph{et al.} reported a population of small RNAs derived from the retrotransposon family GilT/Genie1 located in telomeres, and hypothesized that these may have a role in transposon-silencing \cite{Ullu2005}.

\subsubsection{Unexpected Low Heterozygosity in a Polyploid Organism}
The WB isolate of \emph{G. intestinalis} contained \textless 0.01\% heterozygosity as estimated from genomic reads \cite{Morrison2007}. Asexual organisms with diploid or higher genome ploidy would be expected to accumulate extensive genomic heterozygosity, i.e. single nucleotide polymorphisms between homologous nucleotide sites. The phenomenon is known as the Meselson effect and has been observed in bdelloid rotifers \cite{Welch2000}. However, not all asexual organisms display high levels of heterozygosity. One prominent example is asexual lineages of \emph{Daphnia}, which reduce heterozygosity via ameiotic recombination \cite{Omilian2006}.

Poxleitner \emph{et al.} used an episomal plasmid to demonstrate genetic exchange between nuclei of cysts; a process named diplomixis \cite{Poxleitner2008}, and may partially explain how the parasite maintains low heterozygosity. Carpenter \emph{et al.} showed that cyst formation occurs from a single trophozoite and not by fusion of two trophozoites \cite{Carpenter2012}. The authors of the former study also concluded that nuclear sorting, i.e. each daughter cell receives a pair of identical nuclei, is not likely to be a mechanism by which \emph{G. intestinalis} reduces heterozygosity. The precise mechanism likely involves gene conversion and/or homologous recombination and is yet-to-be described.

\subsubsection{Promiscuous Transcription due to Loose Transcriptional Regulation?}
\emph{G. intestinalis} contains two nuclei with an equal amount of DNA, which has been shown by DAPI staining \cite{Kabnick1990}. Uridine incorporation into RNA showed that both nuclei are transcriptionally active \cite{Kabnick1990}.  Recent data indicate that the two nuclei may not be completely identical: (\emph{i}) Benchimol \emph{et al.} showed that the two nuclei differ in nuclear pore number and distribution \cite{Benchimol2004}; (\emph{ii}) T\r{u}mová \emph{et al.} reported that the nuclei differ in both number and size of chromosomes \cite{Tumova2007}; (\emph{iii}) a microRNA precursor was found in only one nucleus \cite{Saraiya2011}; and (\emph{iv}) Yang \emph{et al.} reported allele-specific expression of one \emph{vsp} \cite{Yang1994}.

Compared with other well-studied eukaryotes, the transcriptional apparatus is simple: 21 of 28 of the eukaryotic RNA polymerase polypeptides are present, but only 4 of the 12 general transcription initiation factors \cite{Best2004}. Sequencing of cDNA clones has found short 5$'$ and 3$'$ untranslated regions, sometimes only a few nucleotides \cite{Adam2001}. Transcriptome profiling using Serial Analysis of Gene Expression (SAGE) and microarrays have uncovered a limited set of differentially expressed genes \cite{Birkeland2010,Morf2010}. Current data on transcription in \emph{G. intestinalis} indicate loose regulation at the transcriptional level. Only a few regulatory promoter elements have been discovered, mainly for developmental genes \cite{Knodler1999,Sun2002}. Intergenic distances are short (the median is 103 bp) \cite{Morrison2007}, leaving little space for regulatory elements. Analysis of promoters has failed to reveal shared motifs or regulatory elements, and suggested that promoters are degenerate \cite{Yee1992,Holberton1995,Morrison2007}. An AT-rich sequence of 8 bp was found to be sufficient to drive transcription \cite{Elmendorf2001}. Hence, AT-richness appears to be the only prerequisite to initiate transcription, possibly explaining the abundance of pervasive transcription in this organism \cite{Teodorovic2007}. One consequence of this organization is bidirectional promoters, which contribute to pervasive transcription \cite{Teodorovic2007}.

Drosha and Exportin-5 are two essential components of the microRNA-processing pathway, both of which are missing in \emph{G. intestinalis} \cite{Saraiya2008}. However, the parasite has Dicer and Argonaut homologs. \emph{G. intestinalis} Dicer has been cloned and shown to produce RNA fragments between 25 to 27 nucleotides \emph{in vitro} \cite{Macrae2006}, although with lower affinity for its small RNA products compared with the human homolog. A recent study used antisense-ribozyme RNA in giardiavirus-infected trophozoites to knockdown expression of the mRNA encoding the Argonaut protein \cite{Saraiya2008}. The authors found that knockdown of Argonaut mRNA inhibited trophozoite replication, and concluded that Argonaut has an important role in the parasite. Moreover, the same study found a snoRNA-derived small RNA of 26-nt length produced by Dicer, and localized it to the cytoplasm. Target sites of these small RNAs were identified in \emph{vsp} genes. An independent study similarly reported \emph{vsp} regulation by RNA interference \cite{Prucca2008}. Altogether, the RNA interference (RNAi) apparatus seems to not be completely analogous with that found in metazoans, which is also supported by the fact that giardiavirus (a double-stranded RNA virus) can replicate in certain strains of the parasite \cite{Wang1993}.

\newpage

\subsection{\emph{Trypanosoma cruzi} -- A Pathogen Transmitted by Blood-sucking Insects and Cause of Systemic Illness}
\fancyhead[RO,LE]{\emph{Trypanosoma cruzi}}

\emph{Trypanosoma cruzi} (\emph{T. cruzi}) is a protozoan parasite and causative agent of Chagas disease (American trypanosomiasis), both discovered and described by Carlos Chagas in 1909 \cite{Chagas1909}. Chagas disease is a zoonosis, affecting {\raise.17ex\hbox{$\scriptstyle\sim$}}8 million people mainly in rural and peri-urban areas of Mexico, Central America and South America \cite{Tarleton2007}. A wide range of insect vectors facilitate transmission of \emph{T. cruzi}, and the endemic range of both the parasite and its vectors stretches from southern United States to Argentinean Patagonia \cite{Rassi2010}. Chagas disease also occurs in non-endemic countries, because of migratory influx from endemic countries in Latin America \cite{Schmunis2007}. More than 300,000 individuals are currently estimated to carry the infection in the United States and \textgreater 80,000 in Europe \cite{Rassi2010}. However, since the natural vectors are not present, the disease is mainly confined to the infected individuals and accidental transmission via blood transfusion or organ transplant.

Chagas disease is a chronic and systemic illness \cite{Rassi2010}. The parasite has likely existed among animals in the Americas for millions of years, as concluded from its wide geographical distribution and host range \cite{Miles2009}. Another line of evidence comes from observations of pathology, where domestic animals and humans often display pathology from the infection, in contrast to wild animals where pathology has not been recorded. This suggests that wild animals and the parasite co-evolved, which led to attenuated virulence. Recovered \emph{T. cruzi} DNA from 9000-year-old mummies indicates that the disease has been troubling humans for extensive time \cite{Aufderheide2004}. In humans, the acute phase of Chagas disease is often asymptomatic and lasts from weeks to months. If symptoms do occur during this phase, they are benign (fever, swollen lymph glands and occasionally, local inflammatory reaction at the bite site) \cite{Rassi2010}. During the acute phase, \emph{T. cruzi} can infect any nucleated cell of the host and parasites may be found in the blood. The immune system eventually reduces the parasitaemia, but does not clear the infection completely and the individual can remain asymptomatic for years or decades. At the chronic stage, parasites are still present in specific tissues, for example, muscle or enteric ganglia. Several years after entering the chronic stage, 20-30\% of the individuals develop irreversible lesions of the heart, colon and/or oesophagus, and in some cases the peripheral nervous system \cite{Rassi2010}. The main lesions of Chagas disease are focal and extensive myocardial fibrosis, driven by a latent inflammatory response. Cardiomyopathy is manifested by cardiac arrhythmias, apical aneurysm, congestive heart failure, thromboembolism and sudden cardiac arrest \cite{Rassi2010}. Chagasic pathology of colon and oesophagus is referred to as the digestive form of the disease, and is caused by destruction of enteric ganglia. The end result is segmental paralysis of the colon and/or oesophagus. Digestive Chagas appears to be more prevalent south of the Amazon basin (Argentina, Brazil, Bolivia and Chile) \cite{Miles2009}.

Treatment of Chagas disease is currently limited to two drugs introduced over 40 years ago, nifurtimox and benznidazole \cite{Rassi2010}. The drugs appear to have variable efficacy, require long treatment periods (60 to 90 days) and do not have an effect towards advanced Chagas disease. The drugs can give rise to severe side effects, including kidney and liver failure. Nifurtimox can also cause neurological disturbances and seizures. Despite many promising new drug targets and leads, for example posaconazole \cite{Pinazo2010}, few candidates have moved beyond the discovery phase -- likely due to limited funding from companies and governments.

\subsubsection{\emph{T. cruzi} is Transmitted by a Wide Range of Insect Vectors}
The parasite, \emph{T. cruzi}, belongs to the kinetoplastid group, which also includes the human parasites \emph{Leishmania} spp. and \emph{T. brucei}. \emph{T. brucei} is indigenous to Africa and \emph{Leishmania} spp. can be found worldwide. Kinetoplastid parasites exhibit some unusual molecular processes, such as RNA editing \cite{Simpson1989}, \emph{trans}-splicing \cite{Campbell2003} and antigenic variation \cite{Barry2001}. \emph{T. cruzi} is transmitted by several different species of insect vectors, mainly of the genera \emph{Triatoma}, \emph{Panstrongylus} and \emph{Rhodnius} (\emph{Hemiptera}; \emph{Reduviidae}). The first entomological description of a Triatomine, \emph{Triatoma rubrofasciata}, was performed already in 1773 by the scientist De Geer \cite{DeGeer1773}. The most important vectors for human transmission are \emph{Triatoma infestans}, \emph{Rhodnius prolixus} and \emph{Triatoma dimidiata} \cite{Rassi2010}. Vector species differ in regional distribution; for example, \emph{T. infestans} has been the most important vector of sub-Amazonian regions, whereas \emph{Rhodnius prolixus} is the predominant vector in northern Latin America. The insects are hematophagous bugs that feed on vertebrate blood, causing transmission of the parasite. Insects often live in cracks of poor quality rural homes or huts, and emerge at night, biting people near the eye or mouth. Many different mammalian hosts act as parasite reservoirs and thereby sustain the transmission cycle of \emph{T. cruzi}. More than 150 species of wild (e.g. armadillos, opossums and raccoons) and domestic (e.g. dogs, cats and guinea pigs) animals can act as \emph{T. cruzi} reservoirs. The disease can also be transmitted via non-vectorial mechanisms, including blood transfusion \cite{Young2007} and organ transplant \cite{CDC2002} as well as congenital transfer from mother to fetus \cite{Gurtler2003,Dorn2007}. Oral transmission is possible via ingested food or liquid \cite{Nobrega2009}, and is generally associated with massive parasite proliferation, with severe and acute clinical manifestations and high rate of mortality \cite{Bastos2010}. However, oral outbreaks are rare but have been documented \cite{ShikanaiYasuda2012}. Even more rarely, individuals have become infected by accidents in the laboratory \cite{Hofflin1987}. Regional differences in disease severity have often been suspected and may be due to parasite genotype, host genetics, transmission cycles and control programs \cite{Miles2009}.

Control measures for Chagas disease include chemical insect control, improvement of housing conditions and education. Blood can be screened before transfusion using serological tests. Most but not all Latin American countries have implemented mandatory serology tests for blood donors. Vector control programs involving spraying of insecticides on houses and buildings, have largely been successful. For example, the ``Southern Cone Initiative'' has reduced Chagas transmission rates in the South Cone of the continent by disrupting transmission via the vector \emph{Triatomina infestans} \cite{Dias2002}.

It is possible that Charles Darwin (1809-1882) contracted Chagas disease during his journey to the Americas, as suggested from descriptions of a specific incident where he was bitten by reduviid insects and from some of the symptoms he suffered later in life \cite{Bernstein1984}.

\subsubsection{The Complex Life Cycle of \emph{T. cruzi}}
\begin{figure}[!ht]
 \centering
 \includegraphics[width=4.5in]{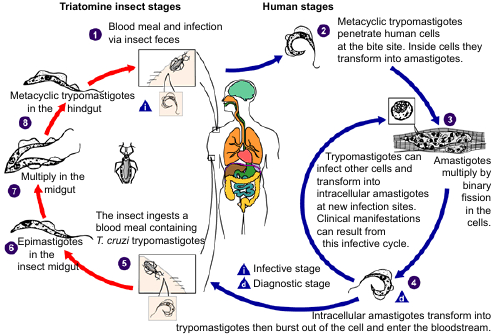} 
  \caption{\small The \emph{T. cruzi} life cycle. Image credit: U.S. Centers of Disease Control and Prevention. }
  
\label{tc_life_cycle}
\end{figure}

\emph{T. cruzi} has a relatively complicated life cycle, with several distinct morphological stages, vector species and mammalian hosts (Figure \ref{tc_life_cycle}). The life cycle begins in a \emph{T. cruzi} reservoir, which is an infected animal or human \cite{Tyler2001}. Infected animals or humans have circulating parasites in the bloodstream. Reduviid insects consume a blood meal from the mammalian reservoir, taking up a population of \emph{T. cruzi} trypomastigotes. Inside the insect, trypomastigotes pass into the midgut and differentiate to amastigote forms. Amastigotes are 3-5 $\mu$m in diameter, proliferate and transform into epimastigotes in the midgut of the insect. There is no evidence that the parasite is harmful to the insect. Epimastigotes also proliferate, and move to the hindgut, where they transform to metacyclic trypomastigotes. Metacyclogenesis may be triggered by substrate interaction of the flagella \cite{Bonaldo1988}. Metacyclic trypomastigotes are then excreted via insect feces, and infection can occur if feces come into contact with the bite wound or mucosal membranes. \emph{T. cruzi} uses its flagella to move into the mammalian host.

\FloatBarrier

Parasites then invade host cells via a mechanism involving the cytoskeleton and host cell lysosomes \cite{Tardieux1992,Rodriguez1996}. Parasites are taken up by the lysosomes, and subsequent acidification inside the lysosomes activates parasite-secreted porin-like molecules that facilitate escape from the vacuole \cite{Andrews1993}. Once inside the cytoplasm, parasites differentiate into amastigotes and begin to proliferate, forming ``pseudocysts,'' and eventually turn into trypomastigotes again. Pseudocysts burst due to the parasite load, and large amounts of parasites are released to the bloodstream and they can infect new cells or get ingested by reduviid insects.

\subsubsection{The Population Structure of \emph{T. cruzi} is Wide, Complex and Contains Signatures of Ancestral Hybridization}
The parasite causing Chagas disease, \emph{Trypanosoma cruzi} \emph{sensu stricto} (\emph{s.s.}), is the type species of the subgenus \emph{Schizotrypanum}. In addition to \emph{T. cruzi s.s.}, the \emph{Schizotrypanum} subgenus harbors approximately half a dozen other trypanosome species, often referred to as \emph{T. cruzi}-like species. Most \emph{T. cruzi}-like species are restricted to bats (order \emph{Chiroptera}), and are morphologically difficult to discriminate \cite{Hoare1972}. One of these bat-restricted organisms is \emph{Trypanosoma cruzi marinkellei} (\emph{T. c. marinkellei}), which was first characterized by Baker \emph{et al.} in 1978 \cite{Baker1978}. \emph{T. c. marinkellei} is regarded as a subspecies of \emph{T. cruzi} and is indigenous to South- and Central American bats. The human infective lineage, \emph{T. cruzi s.s.}, should therefore be referred to as the nominate subspecies \emph{Trypanosoma cruzi cruzi}. However, in this thesis the human infective parasite is simply referred to as \emph{T. cruzi}, or \emph{T. cruzi s.s.} when applicable. Lewis \emph{et al.} estimated the divergence of \emph{T. cruzi s.s.} and \emph{T. c. marinkellei} at 6.51 million years ago using the \emph{gpi} gene \cite{Lewis2011}.

\emph{T. cruzi} propagates predominantly via binary fission. However, Gaunt \emph{et al.} created hybrid clones of distinct \emph{T. cruzi s.s.} strains (\emph{in vitro}), and thereby showed that \emph{T. cruzi s.s.} has an extant capacity for genetic exchange \cite{Gaunt2003}. The mechanism of genetic exchange is somewhat unusual, involving fusion of cells followed by genomic erosion to a diploid genome. Sexual events have likely shaped the current population structure of \emph{T. cruzi s.s.}, but have been sufficiently rare to allow clonal propagation during long periods of time \cite{Brisse2003}. \emph{T. cruzi s.s.} is currently partitioned into six discrete typing units (DTU; TcI-TcVI; Figure \ref{tc_tree2}). Two of these, TcV and TcVI are the result of ancestral hybridization events from TcII and TcIII \cite{Zingales2012}. In addition to the six DTUs, there is one genotype identified only in Brazilian bats, TcBat \cite{Marcili2009}. The genetic heterogeneity of \emph{T. cruzi s.s.} may explain differential clinical manifestations. The null hypothesis of neutral DTU subdivision with respect to Chagas disease severity can safely be rejected, but there is currently no definite correlation between DTU and disease outcome. 

\subsubsection{Was the Ancestor of \emph{T. cruzi sensu stricto} a Bat Trypanosome?}
\begin{figure}[!ht]
 \centering
 \includegraphics[width=5in]{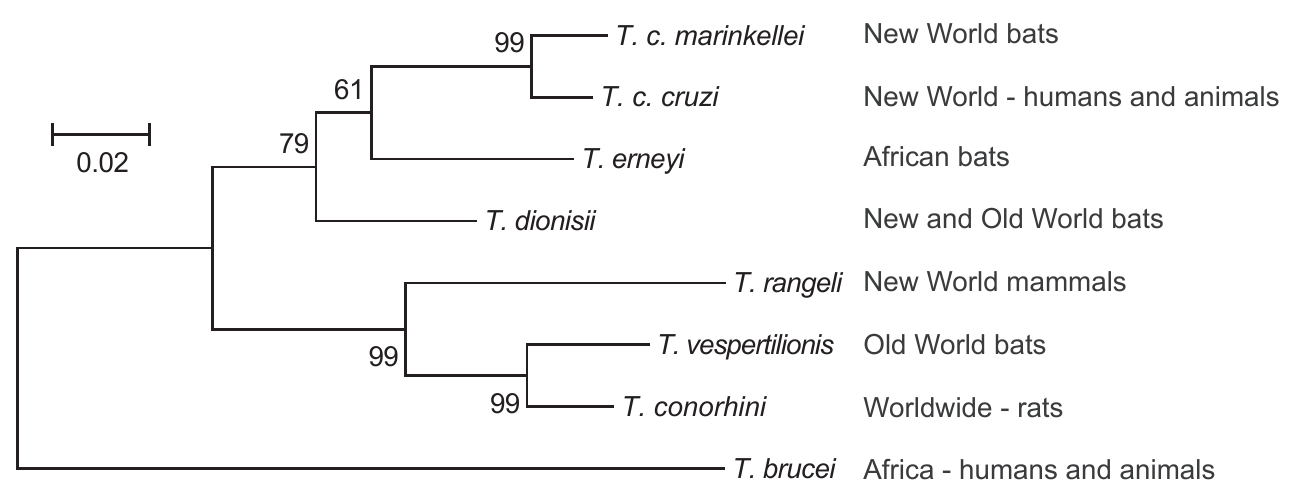}
  \caption{\small Maximum likelihood phylogeny of trypanosomatid species based on the \emph{gGAPDH} gene. Alignments were done with ClustalW2 and inferred with MEGA5 \cite{Tamura2011} using the Tamura-Nei model, 1000 bootstrap replicates were performed. The scale bar refers to number of substitutions per site. Bootstrap values are close to the branches. \emph{T. brucei} was used as an outgroup to support the phylogeny. Geography and hosts are indicated to the right. Accession numbers of the included sequences; JN040964, GQ140362, GQ140360, GQ140358, GQ140364, AJ620283, AJ620267, Tb927.6.4280 (GeneDB).}
  
\label{tryp_species}
\end{figure}

South American trypanosomes diverged from African trypanosomes after the break-up of Gondwanaland, and evolved parasitism independently \cite{Lake1988,Fernandes1993,Maslov1996}. While it is impossible to precisely date the divergence, estimates based on ribosomal RNA genes and biogeographical data, suggest that it occurred 90 to 100 million years from present \cite{Fernandes1993,Stevens1999}. This implies that the divergence of the present day \emph{T. cruzi} and \emph{T. brucei} predated the origins of insect vectors and placental mammals. A phylogeny of the closest known relatives of \emph{T. cruzi} is shown in Figure \ref{tryp_species}. \emph{T. brucei} occurs exclusively in Africa and is transmitted by tsetse flies. In contrast to South American trypanosomes, \emph{T. brucei} could have co-evolved with primates and hominids for many million years. On the other hand, human presence in the Americas stretches no further than {\raise.17ex\hbox{$\scriptstyle\sim$}}30,000 to 40,000 years from present \cite{Bonatto1997}. Hence, \emph{T. cruzi} can only have been in contact with humans for this period of time. An increase in human agricultural activities {\raise.17ex\hbox{$\scriptstyle\sim$}}10,000 years ago was likely the first contact of the parasite with humans, and at that time most infections were likely to have been accidental. Parasite infections then gradually became more prevalent when human dwellings became infested with the insect as an extension of its natural habitat. Possibly, deforestation and an increase of agriculture facilitated the spread of insect vectors and thereby the disease.

Current data indicate a strong association between bats and \emph{T. cruzi}-like flagellates, suggesting a long period of shared evolutionary history. Thomas \emph{et al.} reported that hematophagous arthropods might act as vectors for the transmission of \emph{T. cruzi}-like species among bats \cite{Thomas2007}. Many of the infected bats are insectivorous, suggesting that bats become infected upon feeding on insect vectors. Hamilton \emph{et al.} proposed the hypothesis that the ancestor of \emph{T. cruzi s.s.} was a bat trypanosome, which made multiple jumps to terrestrial mammals \cite{Hamilton2012}. Thus, the broad mammalian host range of \emph{T. cruzi s.s.} may be a characteristic derived from a bat-restricted trypanosome. The following observations support the ``bat-seeding hypothesis'' of \emph{T. cruzi s.s.}: (\emph{i}) the subgenus \emph{Schizotrypanum} is dominated by bat-associated parasites; (\emph{ii}) the closest known relative of \emph{T. cruzi s.s.} is \emph{T. c. marinkellei}, found only in South- and Central American bats; (\emph{iii}) Lima \emph{et al.} reported a new trypanosomatid species of African bats, \emph{Trypanosoma erneyi}, forming a clade within the \emph{Schizotrypanum} with \emph{T. c. marinkellei} as a sister clade \cite{Lima2012} (Figure \ref{tryp_species}); (\emph{iv}) the present day \emph{T. cruzi s.s.} has been found in bats, albeit at low prevalence \cite{Lisboa2008,MaiadaSilva2009,Cavazzana2010}; (\emph{v}) Marcili \emph{et al.} recently reported a new genotype of \emph{T. cruzi s.s.} (TcBat) that is only found in bats \cite{Marcili2009}, which however little is known about and conclusions about its host specificity may reflect insufficient sampling; (\emph{vi}) compared with \emph{T. brucei}, the present day population structure of \emph{T. cruzi} is wider and more complex, consistent with a dispersion facilitated by bats. One study recently reported new strains of the bat-restricted trypanosome \emph{T. dionisii} in British bats, suggesting natural movement of bats between the Old and New World \cite{Hamilton_Cruickshank2012}.

'Ecological host switching' describes a process where parasites may acquire new hosts or expand its host range without evolving new host utilization capabilities \cite{Brooks2005}. The process of ecological host switching has been proposed as the mechanism by which the ancestral \emph{T. cruzi}-lineage colonized terrestrial mammals \cite{Hamilton2007}.

\subsubsection{An Unusual Amount of Genomic Redundancy}
\emph{T. cruzi} strains exhibit extensive variation in DNA content \cite{Engman1987,Wagner1990,Henriksson2005,Lewis2009}, illustrating the diversity of this species. The genome sequence of the \emph{T. cruzi} clone CL Brener (TcVI) has been determined \cite{ElSayed2005}. The CL Brener clone is highly virulent and was isolated from the blood of mice infected with the parental strain CL \cite{Zingales1997}. The CL strain was originally isolated from \emph{Triatoma infestans} in 1963 \cite{Brener1963}. Several factors  complicated genome finishing and resulted in a genome sequence of lower quality than those of \emph{T. brucei} \cite{Berriman2005} and \emph{Leishmania major} \cite{Ivens2005}, which were sequenced in parallel: (\emph{i}) the CL Brener clone was a genetic hybrid, i.e. it consisted of two 3-4\% diverged haplotypes, referred to as non-Esmeraldo-like and Esmeraldo-like; (\emph{ii}) the genome was enriched with sequence repeats of various types, comprising {\raise.17ex\hbox{$\scriptstyle\sim$}}50\% of the genome; and (\emph{iii}) the karyotype of the CL Brener strain was found to be complex, consisting of at least 80 chromosomes of various sizes \cite{Weatherly2009}. Arner \emph{et al.} realigned the shotgun data from the genome project with the assembly and showed that many genes existed in almost identical copies \cite{Arner2007}. This indicated that copy number variation must have been introduced relatively recently, since alleles have had little time to diverge. Weatherly \emph{et al.} organized contigs and scaffolds from the genome project into longer chromosome-wide sequences \cite{Weatherly2009}, although 30-40\% of the genome is still unresolved. The majority of excluded genes belong to gene families. Most of the sequence repeats that went into the original genome draft \cite{ElSayed2005} were genes encoding surface antigens, such as \emph{trans}-sialidases (TSs), mucins, mucin-associated surface proteins (MASPs), dispersed gene family 1 (DGF-1), GP63 peptidases, and retrotransposons. Many of the repeated genes exist as pseudogenes in the genome. One prominent example is the TS family, containing at least 693 pseudogenes in the draft genome sequence, and likely many more alleles that fell outside the assembly \cite{ElSayed2005}. Many of the surface proteins are glycosylated, and they cover and shield the parasite from the host immune system. Some TS proteins transfer siliac acid from the host to mucins \cite{Frasch2000}, and have been proposed as drug targets \cite{Neres2008}. Minning \emph{et al.} used comparative genomic hybridization to sample multiple independent strains and found widespread copy number variation and whole chromosome aneuploidies \cite{Minning2011}.

Most \emph{T. cruzi} genes are densely packed into polycistronic transcription units (PTU), which are separated by strand switch regions \cite{Campbell2003}. RNA polymerase (pol) II drives transcription in two different directions, resulting in polycistronic pre-mRNAs. The pre-mRNA from the PTU is subsequently matured via \emph{trans}-splicing and polyadenylation to mRNAs. \emph{trans}-splicing involves the ligation of a 39-nt spliced-leader sequence to the 5$'$ end of transcripts. Since the life cycle is complex, the parasite needs to regulate its gene expression in order to adapt to different hosts and local environments. While RNA pol II is responsible for the overall transcription of PTUs, the genome lacks defined promoter elements. This organization suggests that individual genes are not regulated at the transcriptional level; rather it is assumed that gene expression is regulated at the post-transcriptional level. However, recent evidence indicates that epigenetic mechanisms may be involved: (\emph{i}) Respuela \emph{et al.} provided evidence of acetylation and methylation at divergent (head to head) strand switch regions, but did not find these patterns at convergent (tail to tail) strand switch regions or within PTUs \cite{Respuela2008}; and (\emph{ii}) Ekanayake \emph{et al.} reported the presence of the glycosylated thymine base ($\beta$-D-glucosyl-hydroxymethyluracil or base J) close to PTUs \cite{Ekanayake2011,Ekanayake_2_2011}. Base J is rare outside of the \emph{Kinetoplastida}, it has only been found in \emph{Diplonema} (a small phagotropic marine flagellate) \cite{vanLeeuwen1998}, and \emph{Euglena gracilis} (a unicellular algae and close relative to kinetplastids) \cite{Dooijes2000}.

\emph{T. cruzi} also contains mitochondrial DNA, which is present in a disk-like structure known as the kinetoplast. Kinetoplast DNA can be divided into maxicircles and minicircles and are circular molecules that are interlocked in a complex network. Their precise size depends on the strain, but maxicircles typically occur in 20-30 copies per cell and range in size from 35 to 50 kb. Minicircles occur in thousands of copies per cell and are approximately 0.8 to 1.6 kb. Transcripts of minicircles and maxicircles are involved in uridine insertion/deletion RNA editing \cite{Landweber1994}. Some evidence indicates that minicircles can integrate into the host genome \cite{SimoesBarbosa2006,Hecht2010}, and therefore have the potential to evoke immune responses and alter host gene expression.

\subsubsection{Lack of RNA interference in \emph{T. cruzi} but not in \emph{T. brucei}}
Small RNAs are non-coding RNA molecules that are either functional or non-functional. In many eukaryotes, functional small RNAs are abundant and partitioned into many different classes, e.g. microRNAs, short interfering RNAs, piwi-interacting RNAs and several others \cite{Brosnan2009}. RNA interference (RNAi) is a gene silencing process, deeply rooted in eukaryotes, mediating silencing via RNA-induced degradation of target transcripts. At the heart of RNAi lies the Argonaute/Piwi protein complex, which exerts post-transcriptional gene silencing. In the kinetoplastids, the presence of RNAi is variable \cite{Lye2010}. Ng\^{o} \emph{et al.} showed already in 1998 that \emph{T. brucei} possesses functional RNAi \cite{Ngo1998,Wang2000}, but RNAi is missing or non-functional in \emph{T. cruzi} \cite{DaRocha2004}. In \emph{Leishmania} spp. the situation is similar, RNAi is present in some species (e.g. \emph{L. braziliensis}, \emph{L. panamensis}, \emph{L. guyanensis}) but not in others (e.g. \emph{L. major}, \emph{L. donovani}, \emph{L. mexicana}) \cite{Lye2010}. These data suggest that RNAi was lost twice in the evolution of the kinetoplastids. Future research will be needed to answer if \emph{T. cruzi}-like organisms are also RNAi-negative. The cause of RNAi loss can only be speculative, but it is possible that loss of active mobile elements freed the parasite from keeping RNAi to mitigate the effects of mobile elements. It is also possible that loss of RNAi was selected for, i.e. if the loss  altered gene expression so that it affected virulence or other properties.

While small RNAs of \emph{T. brucei} have been relatively well studied \cite{Djikeng2001,Wen2011,Michaeli2012}, \emph{T. cruzi} has received less attention on the subject. Analysis of the genome sequence has shown that \emph{T. cruzi} lacks Dicer and Argonaute homologs \cite{ElSayed2005}. However, the genome contains a gene encoding an Argonaute/Piwi-like protein, but apparently without a recognizable PAZ domain \cite{Ullu2004}. The significance of this gene is uncertain. However, the lack of functional RNAi does not exclude the existence of small RNAs that may exert effects through other pathways. Small RNA species of \emph{T. cruzi} containing the spliced leader mini-exon and other small RNAs were  reported early \cite{Hernandez1983,Milhausen1984}. Garcia-Silva \emph{et al.} reported a population of tRNA-derived small RNAs that localized to cytoplasmic granules, and increased during nutritional stress \cite{GarciaSilva2010}.

\begin{center}
\decofourleft\decofourright
\end{center}

\newpage 


\section{Aims}
\fancyhead[RO,LE]{}

\setlength\epigraphwidth{3.0in}
\setlength{\epigraphrule}{0pt} 

\epigraph{\emph{…we have come to the edge of a world of which we have no experience, and where all our preconceptions must be recast.\newline}}{-- D’Arcy Wentworth Thompson (1917), biologist.}

\vspace{1.3cm}

The aim of the thesis was to further understand intraspecific genomic variation and transcriptional features of the protozoan parasites \emph{Giardia intestinalis} and \emph{Trypanosoma cruzi}.

\subsection{Specific aims}

\noindent\textbf{Paper 1}
\newline
Genome sequence comparison of the two human-infecting genotypes of \emph{Giardia intestinalis} (A and B).
\newline

\noindent\textbf{Paper 2}
\newline
Identify genomic features that distinguish a non-human associated genotype of \emph{Giardia intestinalis} (E) from two human-infective genotypes (A and B).
\newline

\noindent\textbf{Paper 3}
\newline
Genome comparison of \emph{Trypanosoma cruzi sensu stricto} with the bat-restricted subspecies \emph{T. cruzi marinkellei}.
\newline

\noindent\textbf{Paper 4}
\newline
Characterization of the short, non-coding transcriptome of \emph{Trypanosoma cruzi} in order to understand if the parasite has functional classes of small RNAs.
\newline

\noindent\textbf{Paper 5}
\newline
Characterization of the transcriptome of \emph{Giardia intestinalis} at single nucleotide resolution and investigation of gene expression divergence.
\newline

\begin{center}
\decofourleft\decofourright
\end{center}

\newpage 


\section{Present investigation}
\fancyhead[RO,LE]{Paper \emph{i} and \emph{ii}}

\setlength\epigraphwidth{4in}
\setlength{\epigraphrule}{0pt} 
\epigraph{\emph{Science is what we have learned about how not to fool ourselves about the way the world is.\newline}}{-- R.P. Feynman (1918 - 1988), theoretical physicist.}

\subsection{Paper \emph{i} and \emph{ii}: Genome Comparison of Three Distinct Isolates of \emph{Giardia intestinalis}}
In Paper \emph{i} and \emph{ii} we performed genome comparisons of three distinct isolates (strains) of \emph{G. intestinalis}, representing genotypes (syn. assemblages) A, B and E (see Figure \ref{giardia_phylogeny} for a phylogeny of genotypes A to G). While A and B infect humans, the E genotype is only associated with hoofed animals. The representative isolates are summarized in Table \ref{tab:isolate_summary}.

\begin{table}[h]
\caption{Summary of compared isolates} 
\centering 
\begin{threeparttable}
\begin{tabular}{c rrrrrrr} 
\hline\hline 
Isolate & Assemblage & Host & Country & Ref. \tnote{a} & Accession \tnote{b} \\ [0.5ex]
\hline 
WB & AI & Human & Afghanistan & \cite{Smith1982} & AACB00000000.1 \tnote{c} \\
GS & B & Human & U.S.A. & \cite{Aggarwal1989} & ACGJ00000000.1 \\
P15 & E & Pig & Czech Rep. & \cite{Koudela1991} & ACVC00000000.1 \\[1ex] 
\hline 
\end{tabular}

\begin{tablenotes}
\footnotesize  \item[a] Original description of the isolate.
  \item[b] NCBI GenBank accession number of the genomic data.
  \item[c] An updated record (AACB00000000.2) is now available.
\end{tablenotes}

\label{tab:isolate_summary}
\end{threeparttable}
\end{table}

Morrison \emph{et al.} described the genome of the WB isolate using Sanger sequencing \cite{Morrison2007}. In paper \emph{i} and \emph{ii} we sequenced the genomes of GS and P15 using Roche/454 sequencing (the former with FLX chemistry and the latter mainly with TIT chemistry; see \cite{Margulies2005} for an overview of the technology). Sequence assembly was performed \emph{de novo} using the assembly software MIRA (Chevreux B, unpublished), and contiguous sequences (contigs) were then improved using targeted Sanger sequencing. Assembly and finishing resulted in 2,931 (N$_{50}$ 34,141 bp) and 820 (N$_{50}$ 71,261 bp) contigs of the GS and P15 genomes respectively. The assemblies were more fragmented than that of WB, but still represented the complete genomes of GS and P15, as confirmed by analysis of non-assembled reads and  assembly sizes (11,001,532 bp of GS and 11,522,052 bp of P15). The P15 assembly was slightly more contiguous, due to the slightly longer read lengths of the TIT chemistry.

Contigs of the two genomes were annotated using a two-tier approach: (\emph{i}) automatic transferring of gene models from the reference strain \cite{Morrison2007}; and (\emph{ii}) followed by manual curation.  The reference genome (WB) was downloaded from the database GiardiaDB \cite{Aurrecoechea2009}, which is part of the EupathDB initiative to integrate genome-wide data sets from eukaryotic pathogens. Open reading frames (ORFs) were extracted from GS and P15, and annotated using best reciprocal BLAST toward genes of WB. Gene models were then manually inspected and unlikely gene models were discarded. Cross-genome comparisons allowed selection of the most conserved and therefore most likely start codon. The majority of the \emph{nek} and \emph{p21.1} families were assigned orthologs, but this was not the case for most \emph{vsp} and \emph{hcmp} genes. This suggested that \emph{vsps} and \emph{hcmps} have undergone lineage-specific diversification events, for example positive selection or recombination.

\begin{figure}[!ht]
  \centering
  \includegraphics[width=4.5in]{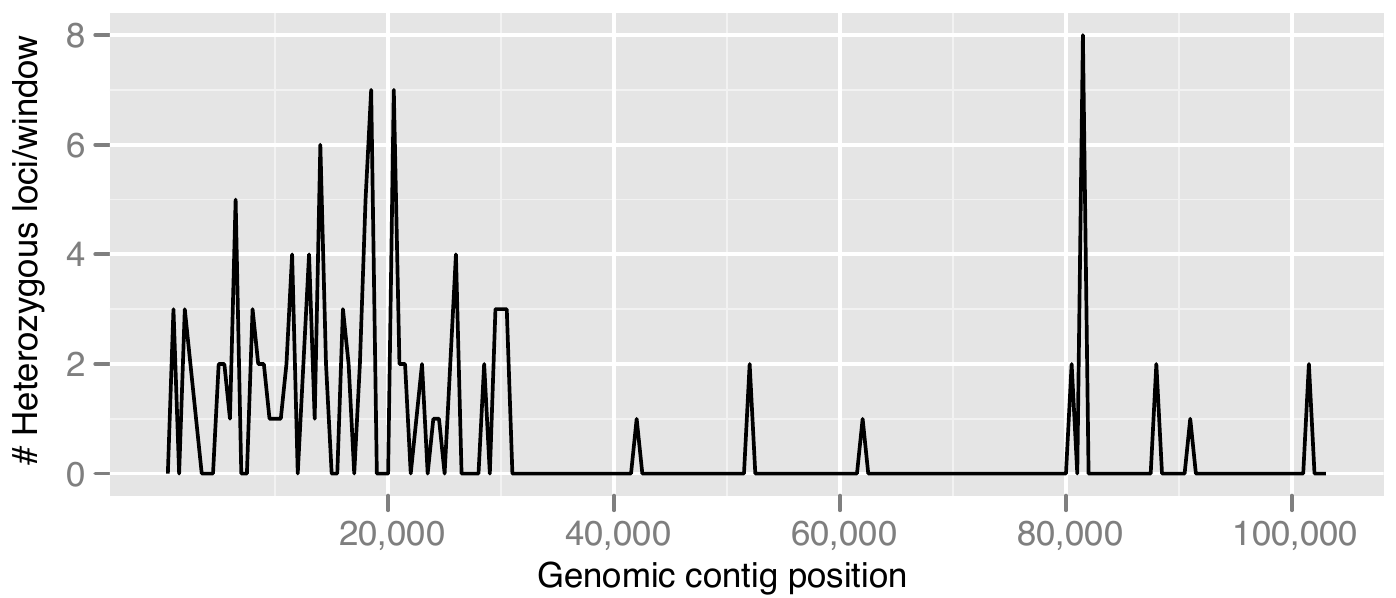}
  \caption{\small Heterozygous loci of the GS isolate counted in sliding windows along a genomic contig. Heterozygous loci were determined using alignments of Roche/454 reads. Windows were of the size 1000 bp and overlapped with 50\%. The contig has the accession number ACGJ01002920. (Y-axis) Heterozygous loci per window. (X-axis) Position along the sequence. }

\label{hz}
\end{figure}
\FloatBarrier

Genomic heterozygosity was almost absent in the WB isolate (\textless 0.01\%) \cite{Morrison2007}, the same was found in P15 with extremely few heterozygous loci. Conversely, GS exhibited extensive genomic heterozygosity, which was genome-wide estimated to {\raise.17ex\hbox{$\scriptstyle\sim$}}0.5\%. The data did not \emph{per se} reveal whether the observed differences were located in the same or different nuclei. Figure \ref{hz} shows how heterozygosity varies along a contig representing 0.83\% of the genome. About half of detected heterozygous loci were located in coding sequences, and 38\% changed amino acid (non-synonymous changes). As expected, there was a strong bias toward transitions. One implication of a  heterozygous genome is the potential to encode additional protein isoforms. Interestingly, heterozygous loci were frequently clustered, intervened by long homozygous regions. The presence of heterozygous loci in clusters rather than homogeneously dispersed over the whole genome are in disfavor of the Meselson effect, i.e. the phenomenon where asexual organisms accumulate heterozygosity in the absence of sexual or ameiotic recombination. A similar mosaic pattern was observed in \emph{Naegleria gruberi} \cite{FritzLaylin2010}. As seen in Figure \ref{hz}, heterozygous loci are more predominant on the left half of this genomic segment. One interpretation of this pattern would be that homogenization via gene conversion has partially taken place. It is also possible that other processes have contributed to heterozygous deficit, such as the Wahlund effect or selfing/homogamy \cite{Tibayrenc2012}. These indirect observations thus suggest that GS has undergone a more recent sexual event as compared with WB and P15. The two latter isolates may have longer asexual histories.

\subsubsection{The Core Genome and Isolate-specific Genes}
The shared and non-shared gene content of the three isolates was investigated using reciprocal BLAST searches. The analysis revealed that the core gene content could be defined by 4557 genes, which excluded isolate-specific genes and \emph{vsps}. The core gene content was comprised of housekeeping genes with homology to other eukaryotes, and \emph{G. intestinalis}-specific genes. Thirty-eight, thirty-one and five genes were specific for P15, GS and WB respectively. One of the P15-specific genes represented an acetyltransferase, and phylogenetic analysis indicated a bacterial origin (likely from a bacterial species of the group Firmicutes). The donor lineage could not be precisely defined, but may be one of \emph{Lactobacillus}, \emph{Cloststridium}, \emph{Anaerotruncus} or \emph{Enterococcus}, all of which are common inhabitants of the gastrointestinal tract. This suggests the uptake of the gene was relatively recent, and is an example of bacteria-to-eukaryote horizontal gene transfer. The GS genome contained several genes that were likely transferred from bacteria, one of which is likely an example of ``dead upon arrival,'' i.e. it was most likely transferred as a pseudogene. At least 96 genes with detectable homology to bacterial genes were conserved in the A, B and E genotypes and have attained housekeeping functions. The mechanism behind horizontal gene transfer is not determined, but likely involves multiple successive steps; where each must be successful in order for the gene to be integrated into the new genome. A successfully integrated gene must also not be deleterious to the new host and convey a selective advantage to become fixed in the population. Frequent exposure to immense bacterial populations in the intestine has most likely contributed to creating opportunities for horizontal gene transfers.

\subsubsection{Structural Variation and Overall Divergence}
Synteny breaks refer to disruption of gene order, often due to genomic rearrangements. In both the GS and P15 (compared to WB), synteny breaks were recorded and tended to occur in regions devoid of housekeeping genes. These regions displayed an atypical nucleotide composition in a sliding-window analysis and deviated in GC-content. Rearrangements may have been introduced spontaneously without affecting parasite fitness, and may therefore have circumvented purifying selection.

The average amino acid identity between WB and P15 was 90\%, 81\% between P15 and GS and 78\% between GS and WB, as measured by comparing orthologous genes. The sequence identities recapitulated that of single-gene phylogenies, confirming the accuracy of previous phylogenetic trees, which are usually not inferred from genome-wide data. The sequence divergence of P15 and WB was similar to what is observed between \emph{L. major} and \emph{L. infantum}, whereas the divergence of GS and WB was similar to that of \emph{Theileria parva} and \emph{T. annulata}. Hence, the divergence between the studied \emph{G. intestinalis} genotypes is similar to what is observed between distinct species. It can thus be argued that the \emph{G. intestinalis} genotypes should be regarded as separate species rather than genotypes.

The dN/dS ratio (rate of non-synonymous nucleotide substitutions/rate of synonymous nucleotide substitutions) can be used to indirectly identify positive selection. The WB vs. GS comparison did not allow calculation of dN/dS ratios since synonymous changes were saturated (i.e. more than one substitution per site). Analysis of dN/dS of WB and P15 indicated, as expected, that most of the genome was under purifying selection. Several uncharacterized genes were found to exhibit elevated dN/dS ratios (\textgreater 1), indicating putative positive selection. Gene Ontology analysis indicated that five GO categories contained genes under positive selection, possibly reflecting co-evolution of multiple genes involved in common pathways. Furthermore, SAGE data were used to categorize genes into developmental categories. Four developmental categories displayed elevated dN/dS ratios, possibly reflecting lineage-specific divergence of cellular processes.

\newpage

\subsection{Paper \emph{iii}: Genome Comparison of \emph{Trypanosoma cruzi sensu stricto} With the Bat-restricted Subspecies \emph{T. cruzi marinkellei}}
\fancyhead[RO,LE]{Paper \emph{iii}}

\begin{figure}[!ht]
 \centering
 \includegraphics[width=4.5in]{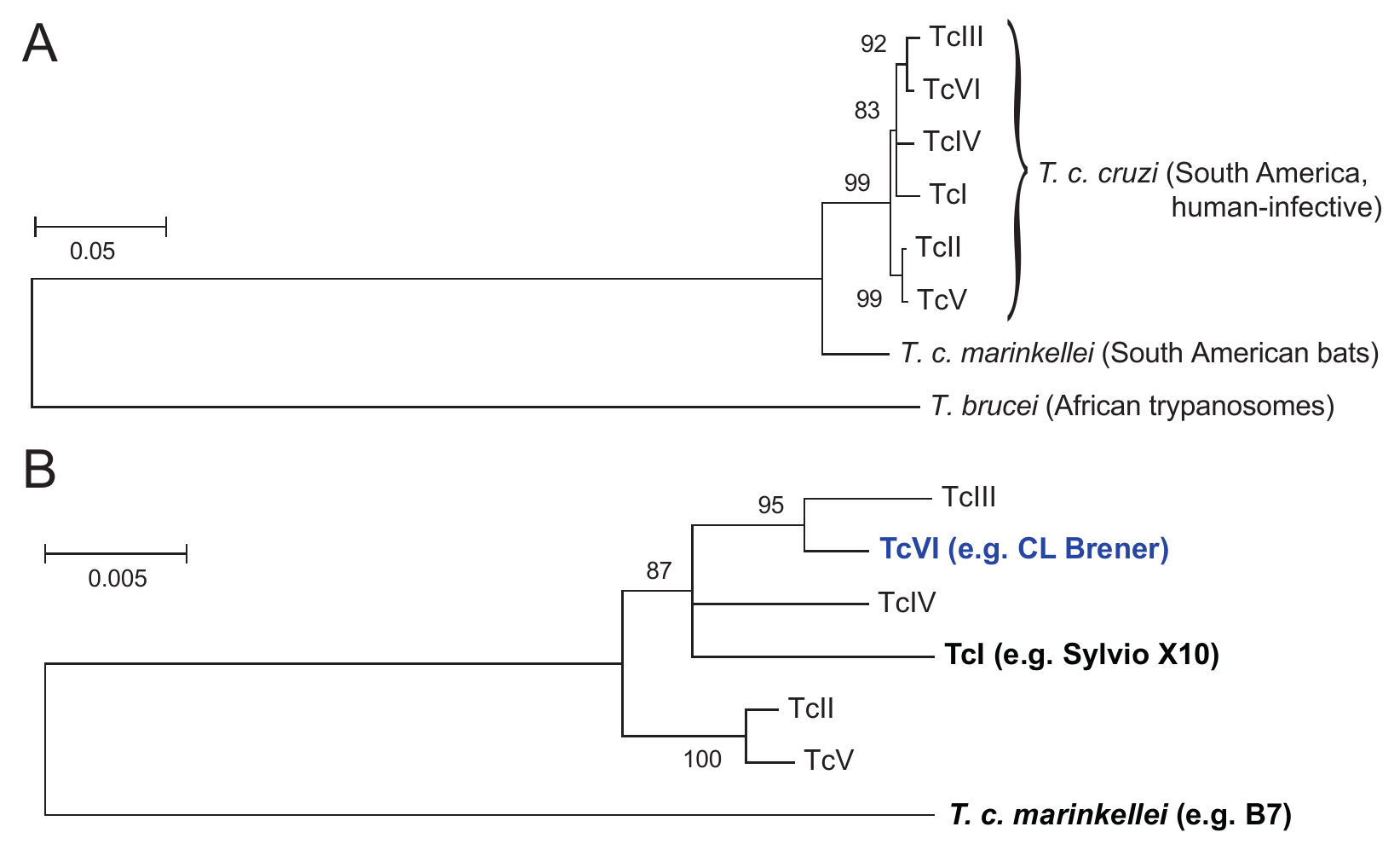}
  \caption{ \small Nucleotide maximum likelihood phylogenies of \emph{T. cruzi s.s.} DTU TcI-VI. The phylogeny was inferred from concatenated sequences encoding the beta-adaptin and endomembrane proteins. Sequences were aligned with ClustalW2 and inferred with MEGA5 \cite{Tamura2011} using the Tamura-Nei model. 1000 bootstrap replicates were performed. Scale bars refer to number of substitutions per site. Numbers close to branches indicate bootstrap support values. \emph{T. brucei} (A) and \emph{T. c. marinkellei} (B) were used as outgroups. Tip labels in bold indicate genotypes sequenced in the present study. (Blue tip label) Available reference genome (non-Esmeraldo-like haplotype). (Accession numbers) of the dataset: HQ859539, HQ859587, HQ859534, HQ859592, HQ859540, HQ859582, HQ859535, HQ859583, HQ859538, HQ859590, HQ859543, HQ859585, HQ859541, HQ859593, Tb927.10.8040, Tb11.02.0960. }
  
\label{tc_tree2}
\end{figure}

The genome of the human infective \emph{T. cruzi} (\emph{T. c. cruzi}; \emph{T. cruzi s.s.}) was compared with that of its closest relative, \emph{T. c. marinkellei} (\emph{Tcm}). Two clones were selected for the comparison: (\emph{i}) \emph{T. cruzi s.s.} Sylvio X10, which was isolated in 1983 from a human male in Pará State Brazil, and has confirmed pathogenicity \cite{Postan1983}. Sylvio X10 is subgrouped into DTU TcI. (\emph{ii}) \emph{Tcm} clone B7, which was originally isolated in 1974 from the bat host \emph{Phyllostomus discolor} in S\~{a}o Felipe Bahia Brazil (M.A. Miles and T.V. Barrett) \cite{Baker1978}. \emph{Tcm} B7 was not found to be infective in immunocompromised mice, nor did it provide immunological protection against subsequent challenge with \emph{T. cruzi s.s.}, suggesting distinct antigenic profiles \cite{Baker1978}. \emph{Tcm} is restricted to South American bats and has to date not been recovered from humans. The phylogenetic relationship of \emph{T. cruzi s.s.} and \emph{Tcm} is shown in Figure \ref{tc_tree2}.

\emph{T. cruzi s.s.} Sylvio X10 (SX10) is a non-hybrid strain, and it has a smaller genome \cite{Lewis2009} than the previously sequenced \emph{T. cruzi s.s.} strain CL Brener (TcVI; Figure \ref{tc_tree2}) \cite{ElSayed2005}. The smaller genome and non-hybrid nature implies that this clone likely has fewer repetitive sequences. \emph{T. cruzi s.s.} SX10 is also evolutionarily distinct to \emph{T. cruzi s.s.} CL Brener, and therefore creates an interesting basis for comparison.

\begin{table}[h]
\caption{Summary of compared \emph{T. cruzi} clones}
\centering
\begin{threeparttable}
\begin{tabular}{c rrrrrrr}
\hline\hline 
Species & Subspecies & DTU \tnote{a} & Clone & Host & Ref. \\ [0.5ex]
\hline 
\emph{T. cruzi} & \emph{T. c. cruzi} & TcI & Sylvio X10/1 & Human & \cite{Postan1983} \\
\emph{T. cruzi} & \emph{T. c. cruzi} & TcVI & CL Brener & Human & \cite{Brener1963} \\
\emph{T. cruzi} & \emph{T. c. marinkellei} & - & B7 & Bat & \cite{Baker1978} \\[1ex] 
\hline 
\end{tabular}

\begin{tablenotes}
\footnotesize \item[a] Discrete Typing Unit; genotype.
\end{tablenotes}

\label{tab:tc_summary}
\end{threeparttable}
\end{table}

The study began with Roche/454 and Illumina sequencing of sheared genomic DNA of \emph{T. cruzi s.s.} SX10 and \emph{Tcm} B7. Briefly, Roche/454 and Illumina reads were assembled separately \emph{de novo} and the assemblies were then merged into a non-redundant assembly. The merged assemblies were subjected to quality enhancements, including scaffolding, gap closure and homopolymere error correction. The \emph{T. cruzi s.s.} CL Brener genome was used for transferring gene models to the new genomes, since the majority of the gene repertoire was expected to be shared. The genomes were annotated using a semi-automatic pipeline. Orthologs were identified using best reciprocal BLAST, and gene models were manually curated. Flow cytometry estimated the genome sizes of \emph{T. cruzi s.s.} SX10 and \emph{Tcm} B7 to 44 and 39 Mb respectively (haploid genome sizes). The haploid genome size of \emph{T. cruzi s.s.} CL Brener has previously been estimated to 55 Mb \cite{ElSayed2005}. The assembly sizes of the genomes closely correlated with the experimental measures, which confirmed the computational steps.

\FloatBarrier

Heterozygosity in the \emph{T. cruzi s.s.} SX10 and \emph{Tcm} B7 was estimated to 0.19 and 0.22\% respectively. Sliding window analysis revealed that heterozygous sites were often clustered in blocks. The organization of heterozygous loci therefore resembled that found in \emph{T. cruzi s.s.} CL Brener, although at much lower levels. One could speculate that the mosaic structure is a result of gene conversion. Overall, the heterozygosity levels were similar to those found in some \emph{Leishmania} species \cite{Rogers2011}.

\subsubsection{Evidence of a Recent Eukaryote-to-Eukaryote Horizontal Gene Transfer}
The genomes of \emph{T. cruzi s.s.} SX10, \emph{T. cruzi s.s.} CL Brener and \emph{Tcm} B7 were systematically searched for gene differences. The search identified a unique 1,662 bp acetyltransferase gene in \emph{Tcm} B7. The gene will be referred to by its locus tag MOQ\_006101. Sequence analysis revealed that the flanking genomic regions were present in \emph{T. cruzi s.s.} SX10 and \emph{T. cruzi s.s.} CL Brener, but not the gene itself. MOQ\_006101 was not identified in unassembled genomic reads of \emph{T. cruzi s.s.} SX10 or \emph{T. cruzi s.s.} CL Brener, suggesting it is unique to \emph{Tcm} B7. Several fragments of VIPER retrotransposons were identified close to the locus, and RT-qPCR confirmed expression of the gene.

Phylogenetic reconstruction indicated that the closest known homologs were in algae and plants, suggesting MOQ\_006101 was transferred from another eukaryote. MOQ\_006101 showed low sequence identity (30-50\%) toward genes in NCBI GenBank. The absence of intron-exon boundaries suggested MOQ\_006101 was transferred as a spliced mRNA, likely from a species not represented in GenBank.

The GC-content of the gene was compared with the global GC-content in coding sequences. GC-content analysis indicated an unusual composition compared to the rest of the genome, strengthening the notion of a transfer from another species. It remains to be determined whether MOQ\_006101 encodes a functional protein. Moreover, examination of multiple \emph{Tcm} isolates may answer whether MOQ\_006101 has been fixed in the species or only in a certain lineage. Whether functional or not, the gene itself is interesting since it represents an unusual instance of horizontal gene transfer between two eukaryotes.

\subsubsection{\emph{Trypanosoma cruzi marinkellei} has Capacity to Invade non-Bat Epithelial Cells}
The potential of \emph{Tcm} B7 to invade cell lines other than bat was investigated. Three common cell lines were selected: (\emph{i}) Vero cells (kidney cells from African green monkey); (\emph{ii}) OK cells (from a North American opossum); and (\emph{iii}) Tb1-lu cells (bat lung). The experiments showed that \emph{Tcm} B7 has retained the capacity to invade each of the three cell lines, despite that the cells were originally derived from different species. Surprisingly, there was no preference for bat epithelial cells. Prolonged incubation of the parasite with cells showed that \emph{Tcm} B7 is able to replicate intracellularly and the invasion process appears to be analogous to \emph{T. cruzi s.s.}. In conclusion, these data indicate that the bat-specificity of \emph{Tcm} is not mediated at the cell-invasion level.

\newpage

\subsection{Paper \emph{iv}: The Small RNA Component of the \emph{Trypanosoma cruzi} Transcriptome}
\fancyhead[RO,LE]{Paper \emph{iv}}

A cDNA library of small RNAs (sRNAs) of the size range 16 to 61 nucleotides (nt) was prepared and sequenced from \emph{T. cruzi} CL Brener epimastigotes, i.e. the insect stage of the parasite. Epimastigotes were used due to the ease of obtaining sufficient RNA from this life stage. Other stages often require cultivation together with mammalian cells, which may contaminate the subsequent RNA preparation. We also assumed that functional sRNAs, if present at all, would also be present in epimastigotes. The particular size range of 16 to 61 nucleotides was selected to avoid spliced leader RNA, which could otherwise cloud the analysis.

Sequencing generated 582,243 sRNAs, of which 90.7\% aligned with the genome sequence. We subsequently used the annotation of the genome to assign sRNAs into relevant categories, i.e. if an sRNA overlapped a tRNA, it was assigned to the tRNA category, etc. With respect to the non-Esmeraldo-like haplotype, 72.1\% of the sRNAs derived from transfer RNAs (tRNA-derived small RNAs; tsRNAs).  97.4\% of sRNAs were derived from only three classes of non-coding RNAs (transfer RNA, ribosomal RNA and small nuclear RNA). Only 0.18\% of sRNAs were derived from protein-coding genes. 2.42\% of sRNAs could not be grouped into any canonical RNA class. A few of the small RNAs were experimentally validated using a real time quantitative PCR-based assay. In conclusion, the bulk of sRNAs of the 16 to 61 nt size range were derived from known classes of non-coding RNAs.

\begin{figure}[!ht]
 \centering
 \includegraphics[width=3in]{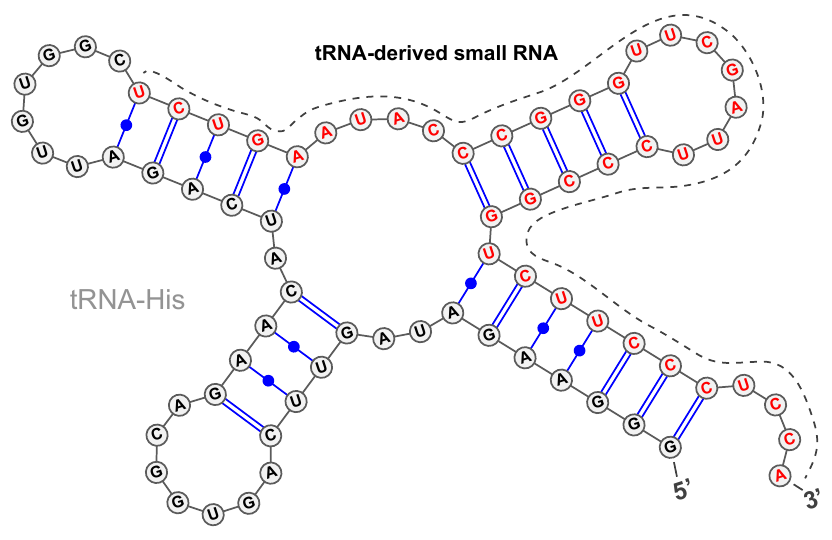}
  \caption{\small Secondary structure of tRNA-His from \emph{T. cruzi} CL Brener (Tc00.1047053508861.10). The secondary structure was predicted using tRNAScan-SE \cite{Lowe1997} and visualized using VARNA \cite{Darty2009}. The sequence in red indicates the part cleaved into a small RNA with copy number 41,929 in the data set.}
  
\label{trna_secondary_structure}
\end{figure}

The median length of tsRNAs was 38 nt, and 88.9\% of them were derived from the tRNA 3$'$ end. One example of a tRNA-derived small RNA is shown in Figure \ref{trna_secondary_structure}. 75.3\% of the 3$'$-derived tsRNAs contained the post-transcriptional `CCA' extension, a hallmark of mature tRNAs. This indicated that most tsRNAs were derived from mature tRNAs, albeit not all. However, it is possible that the `CCA' extension has been lost during sample or library preparation. If tsRNAs would represent degradation products of tRNA turnover, one would expect a correlation between sRNA copy number and the expression level of tRNA genes. Analyses of amino acid usage as a substitute of tRNA gene expression data did not find any correlation. The cleavage site of tRNAs was present inside the anticodon loop, suggesting endonucleolytic cleavage as the responsible mechanism of generation.

\FloatBarrier

1.69\% of the small RNAs were not derived from known non-coding RNAs. These small RNAs were clustered based on their genomic alignment coordinates. Clustering formed 92 distinct expression loci, of which homology searches revealed known non-coding RNAs for 13 loci. The remaining 79 loci did not fall into known non-coding RNA classes and had an average length of 54 nt. No homology was found in Rfam or GenBank databases. 35 of the novel RNAs folded into non-hairpin secondary structures and 18 folded into hairpin structures. 1,159 small RNAs were not clustered and had a median length of 24 nt. Of these small RNAs, BLAST searches revealed 335 to be derived from rRNA and 819 from protein-coding genes. MicroRNA target site prediction of the latter population using the ``seed region'' (nt 2-8) predicted the possibility that these regulate certain categories of mammalian genes. It is therefore tempting to speculate that the parasite may use small RNAs for inter-cell communication, or possibly modification of the gene expression response of the host cell.

0.13\% of small RNAs were derived from repeats, including retroelements. In particular the CZAR element contained 446 mapped small RNAs, which may indicate putative initiation fragments from the transcription of these elements. There was no overrepresentation of antisense reads in any repeat class, suggesting that small RNAs have no role in perturbation of mobile elements. Searches did not reveal any canonical classes of regulatory sRNAs found in metazoans. This is consistent with the lack of RNA interference.

\newpage

\subsection{Paper \emph{v}: Transcriptome Profiling at Single Nucleotide Resolution of Diverged \emph{G. intestinalis} Genotypes Using Paired-end RNA-Seq}
\fancyhead[RO,LE]{Paper \emph{v}}

\begin{figure}[!ht]
 \centering
 \includegraphics[width=4.5in]{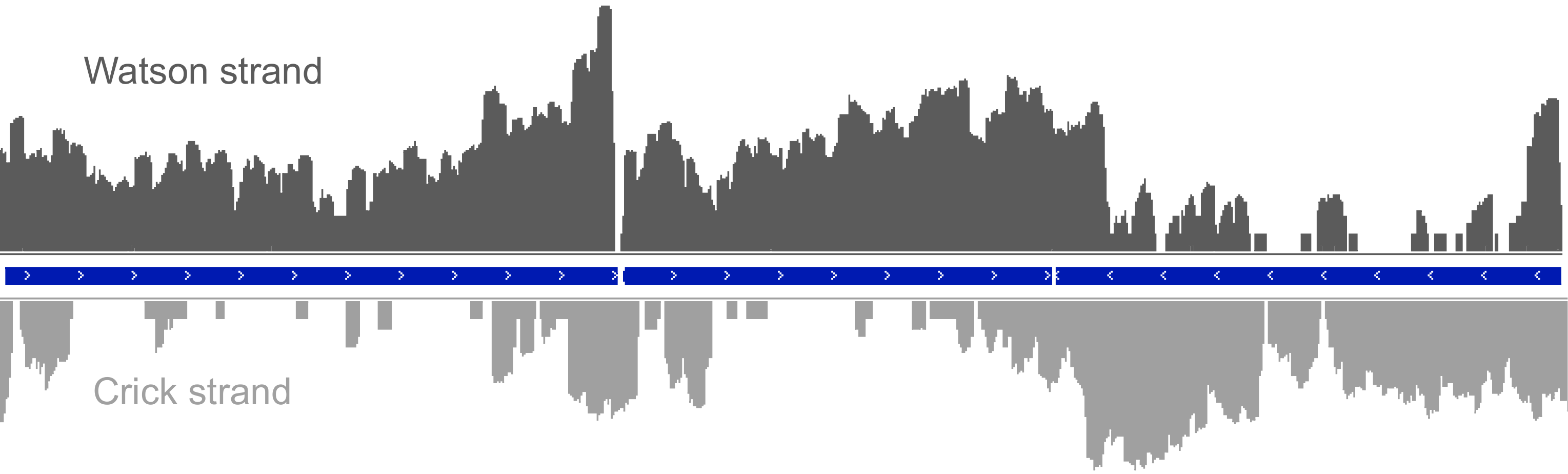}
  \caption{ \small Coverage of RNA-seq sequence reads on three open reading frames (logarithmic scale; blue horizontal bars: GL50803\_92132, GL50803\_92134, GL50803\_40369). (Watson strand; ``plus'') Top. (Crick; ``minus'') Bottom. The strand of the open reading frame is shown by arrows (`\textgreater' refers to the plus strand and `\textless' refers to the minus strand).}
  
\label{cov}
\end{figure}
\FloatBarrier

In this study we performed transcriptome sequencing (RNA-Seq) and comparison of the polyadenylated transcriptomes of four diverged isolates of \emph{G. intestinalis}. The specific aims of the study were to: (\emph{i}) identify genotype-specific patterns of gene expression; (\emph{ii}) confirm and refine genome annotations; and (\emph{iii}) identify qualitative transcript properties like 3$'$ untranslated regions (UTRs). Total polyadenylated RNA was harvested from \emph{in vitro} grown trophozoites of the four isolates WB (AI), AS175 (AII), P15 (E) and GS (B). Sequencing libraries were prepared according to a strand-specific paired-end protocol and sequenced on Illumina HiSeq 2000 as 2x100-nt reads. Each library was sequenced as two technical replicates in order to estimate technical variation. The reproducibility of the RNA-Seq method was determined using two biological replicates of AS175. RNA-Seq generated 33 to 41 million read-pairs from each library, which were aligned (mapped) to the corresponding reference genome (Table \ref{tab:rna_seq_mapped_data}). Figure \ref{cov} shows the strand-specificity of the obtained data. The aligned RNA-Seq data were then used to calculate digital gene expression values, formulated as fragments per kilobase of transcript per million fragments mapped (FPKM). 49 genes were selected for RT-qPCR validation, which confirmed the accuracy of the measurements. Moreover, a global comparison was performed with three microarray data sets from GiardiaDB \cite{Aurrecoechea2009}, indicating a moderate but significant correlation of the two techniques. RNA-Seq measurements did not correlate with SAGE data, which is not surprising since SAGE is generally not quantitative.


\begin{table}[h]
\caption{Summary of studied isolates and generated RNA-Seq data} 
\centering 
\begin{threeparttable}
\begin{tabular}{c rrrrrrr} 
\hline\hline 
Isolate & Assemblage & Ref. & \#Mapped \tnote{b} & \%ORFs \tnote{c} & \%Orthologs \tnote{d} \\ [0.5ex]
\hline 
WB & AI & \cite{Morrison2007} & 37,888,422 & 93.7 & 99.5 \\ 
AS175 & AII & \tnote{a} & 36,878,839 & 98.3 & 99.8 \\
P15 & E & \cite{JerlstromHultqvist2010} & 36,437,138 & 96.3 & 99.6 \\
GS & B & \cite{Franzen2009} & 31,869,806 & 97.3 & 99.7 \\[1ex] 
\hline 
\end{tabular}

\begin{tablenotes}
  \footnotesize \item[a] The genome sequence is not published.
  \item[b] Reads of this strain that uniquely mapped to the reference genome.
  \item[c] Percentage of [annotated] ORFs with detectable transcription (FPKM\textgreater 0.5).
  \item[d] Percentage of conserved four-way orthologs with detectable transcription.
\end{tablenotes}

\label{tab:rna_seq_mapped_data}
\end{threeparttable}
\end{table}

\subsubsection{Low Biological and Technical Variation}
Technical variation consists of measurement imprecision introduced during library preparation or by the sequencing instrument. The technical replicates of each sequencing library indicated very low technical variation (Pearson's \emph{r}$^{2}$=0.99). In this study technical replicates were subjected to the same library construction procedure, and thus reflected variation of the sequencing instrument. On the contrary, biological replicates reflect both technical and biological variation. Biological variation may result from uncontrolled environmental cues or from stochastic variation in gene expression. Biological variation was low (\emph{r}$^{2}$=0.97), and we therefore concluded that gene expression measurements were reproducible. The number of genes involved in the culture-induced biological response was estimated with a $\chi$$^{2}$ test, and found to be 4\% of the total gene content at \emph{p}=0.01 (AS175). There was no meaningful way to group the implicated genes.

\subsubsection{Gene Expression Levels Recapitulate the Known Phylogeny}
\begin{figure}[!ht]
 \centering
 \includegraphics[width=4.5in]{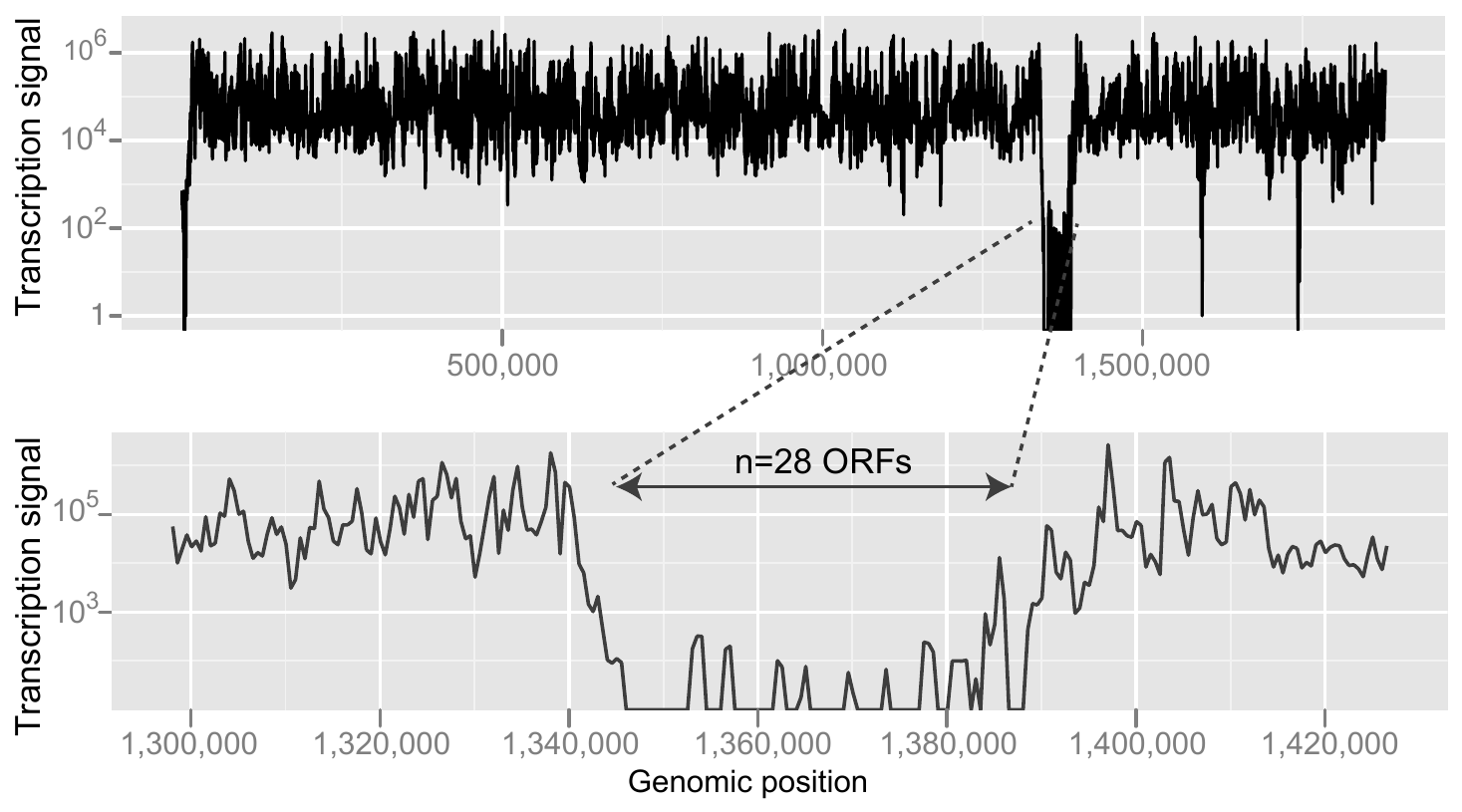}
  \caption{ \small Sliding window analysis of RNA-Seq coverage on scaffold CH991767 (WB isolate). The zoomed in region shows drop in RNA-Seq coverage along a 41 kb region. Sequence coverage was calculated in 500 bp non-overlapping windows. (Y-axis) log$_{10}$-transformed coverage sum. (X-axis) Position along scaffold. }
  
\label{coverage_silent}
\end{figure}

Almost the entire \emph{G. intestinalis} genome was transcribed to some extent, and provided transcriptional evidence for \textgreater 99\% of the conserved genes (Table \ref{tab:rna_seq_mapped_data}). Transcription levels exhibited a wide dynamic range; the fold difference of the median of the 5\% lowest and highest expressed genes was 873X. Notably, a transcriptionally silent gene cluster was identified on chromosome 5 of the WB isolate, encompassing 28 genes in tandem (Figure \ref{coverage_silent}). The genomic region was 41 kb and contained genes associated with replication and genes of no known function. Sequence signatures of these genes were also identified in the GS isolate, suggesting that the region was acquired prior to the split of the lineage leading to the extant A and B genotypes. Because the region exhibited higher than average divergence between A and B, it is possible that it has been subjected to genetic drift without purifying selection. Saraiya \emph{et al.} reported one ORF of this cluster to transcribe a microRNA-like small RNA \cite{Saraiya2011}, suggesting that the region may have certain functionality and provides an explanation to why it has not been lost. Nevertheless, the lack of detectable transcription suggests it may be a selected feature, possibly due to negative effects of parasitic DNA on fitness.

Global gene expression profiles of the isolates were compared and genotype-specific gene expression was identified using a $\chi$$^{2}$ test. A relatively limited number of genes were differentially expressed (31 to 145 genes). These numbers likely represent the lower detection bound since the implemented analysis was conservative. It remains to be investigated how many of the differentially expressed genes are of functional importance.

The theory of neutral evolution states that most nucleotide changes are randomly introduced by neutral drift without affecting fitness \cite{Kimura1968}. On the contrary, positive selection is driven by advantageous mutations and is generally more rare. It is currently accepted that coding sequences (CDS) predominantly evolve by neutral evolution followed by purifying selection \cite{Nielsen2005}. Less is known about the modes of evolution acting on gene expression. We investigated if the rate of gene expression divergence was correlated with that of the CDS. CDS divergence was estimated by the rate of non-synonymous nucleotide substitutions (dN), while gene expression divergence was estimated by FPKM fold change. The analysis was performed genome-wide on the maximum number of defined ortholog pairs (the precise number varied slightly between isolates). Cross-correlations of any two isolates resulted in Pearson's \emph{r} ranging from 0.069 to 0.11, indicating a weak correlation of the two variables. Conversely, there was no correlation between the rate of synonymous nucleotide substitutions (dS) and gene expression divergence (Pearson's \emph{r}=0). Interestingly, highly expressed genes tended to have lower rate of divergence (dN). These data indicated a limited coupling between CDS and gene expression divergence, suggesting that random drift has been the predominant evolutionary mode of gene expression divergence in \emph{G. intestinalis}.

\subsubsection{Promiscuous Polyadenylation Sites and Unusual \emph{cis}-acting Signals}
Polyadenylation sites (polyA sites) were precisely mapped using RNA-Seq data, which allowed global analysis of the \emph{G. intestinalis} `polyadenylation landscape'. PolyA sites were mapped using reads (polyA tags) containing the mRNA:polyA junction (Table \ref{tab:polya_sites}). Aligned polyA tags revealed 22,221 to 49,027 distinct polyA sites (depending on the isolate; Table \ref{tab:polya_sites}).

\begin{table}[h]
\caption{Mapped polyadenylation sites} 
\centering 
\begin{threeparttable}
\begin{tabular}{c rrrrrrr} 
\hline\hline 
Isolate & \#tags \tnote{a} & \#sites \tnote{b} & \#PACs \tnote{c} & \#transcripts \tnote{d} & Median 3$'$ UTR (nt) \\ [0.5ex]
\hline 
WB & 456,928 & 49,027 & 7,617 & 3,884 & 80 \\ 
AS175 & 183,454 & 37,028 & 5,028 & 3,057 & 100 \\
P15 & 436,720 & 51,499 & 8,037 & 3,800 & 83 \\
GS & 71,118 & 22,221 & 2,624 & 1,651 & 85 \\[1ex] 
\hline 
\end{tabular}

\begin{tablenotes}
\footnotesize \item[a] PolyA-tags with mapping quality \textgreater 40.
  \item[b] Unique polyadenylation sites.
  \item[c] PolyA sites within 10 nt were clustered into polyadenylation site clusters. The numbers refer to number of clusters with $\geq$4 tags.
  \item[d] Transcripts with an associated polyA site.
\end{tablenotes}

\label{tab:polya_sites}
\end{threeparttable}
\end{table}

\FloatBarrier

Each of the polyA sites represents the 3$'$ end of an independent transcript, although not necessarily an mRNA. PolyA sites were found to exhibit microheterogeneity, i.e. imprecision of a few nucleotides. Microheterogeneity of polyA sites has been documented in higher eukaryotes and is attributed to the imprecise nature of the polyadenylation machinery \cite{Pauws2001,Wu2011}. The phenomenon is not to be confused with alternative polyadenylation, which is regulated by distinct polyadenylation signals. To account for this heterogeneity, polyA sites within 10 nt were clustered into polyadenylation site clusters (PAC). PACs were assigned to the most likely ORF based on proximity, i.e. the 5$'$-most PAC counted from the translational stop codon was assigned to the ORF. Cloning and 3$'$ rapid amplification of cDNA ends was performed on 9 genes for validation, confirming the accuracy of the mapped sites. The median 3$'$ UTR length was found to be 80 nt for WB, and similar for the other isolates. In comparison, the median 3$'$ UTR length of \emph{Saccharomyces cerevisiae} is 104 nt \cite{Nagalakshmi2008}. This is longer than earlier estimates (around 30 nt) from a small set of highly expressed genes. Several microRNAs have been identified in \emph{G. intestinalis} but searches for target sequences were only done within the first 50 bp from the stop codons. These results suggest that regulation of gene expression via microRNA binding to 3$'$ UTRs can be common since many mRNAs have relatively long 3$'$ UTRs.

The mapped polyA sites were analyzed for putative \emph{cis}-acting signals, i.e. polyadenylation signals (PAS). Positions -40 to -1 of each polyA site were searched for overrepresented hexamers using an iterative algorithm described by Beaudoing \emph{et al.} \cite{Beaudoing2000}. The search identified 13 prominent hexamers, which represent putative PAS. None of these were identical with the canonical eukaryotic PAS (AAUAAA). However, 5 of the 13 putative \emph{G. intestinalis} PAS contained the tetranucleotide 'UAAA', which is a part of the canonical eukaryotic motif. The tetramer was located approximately 10 nt from the polyA site. Interestingly, polyA sites located in the sense orientation tended to have fewer of the identified hexamers in contrast to antisense or intergenic polyA sites.

The nucleotide composition surrounding polyA sites was analyzed, which indicated a distinct pattern of AU-richness. This pattern may be required for recognition by the polyadenylation machinery, or for binding of polyadenylation factors. 81\% of tail-to-tail gene pairs had 3$'$ UTRs that overlapped the transcription unit of an adjacent gene on the opposite strand. This transcriptional organization may be a way of gene regulation, but also causes the production of double stranded RNA. Such pervasive transcription is often not compatible with functional RNA interference. While \emph{G. intestinalis} has some components of RNAi, it is not completely identical to that found in higher eukaryotes. It can therefore be speculated if \emph{G. intestinalis} has lost certain components of RNAi in favour of genomic fidelity and minimalism.

\subsubsection{Biallelic Transcription and Correlation of Allele Expression and Allele Dosage}
As described in paper \emph{i}, the GS isolate contains {\raise.17ex\hbox{$\scriptstyle\sim$}}0.5\% genome-wide heterozygosity. The RNA-Seq data were used to confirm expression of the identified alleles and to study allele-specific expression (ASE). Mapping bias is a major problem in ASE assays, which means that current mapping algorithms preferentially map one of the alleles (discussed in for example \cite{Degner2009}). To further understand the extent of mapping bias, we generated and mapped simulated RNA-Seq data. The simulated data followed realistic error profiles, and did not indicate systematic mapping bias, but nevertheless indicated an inherent bias toward mapping of one allele. The amount of bias was influenced by sequence errors, but likely also other factors.

The simulated data was modeled using a Cauchy distribution and used for significance testing. 98\% of the genes with at least one heterozygous locus displayed biallelic transcription, i.e. the two alleles were identified at the transcriptional level. Of these genes, 82\% indicated allelic expression imbalance at \emph{p}=0.05, i.e. not an equal number of reads were derived from each allele. We examined if the allelic expression ratio corresponded to the observed allele count. Allele counts were inferred from the depth of genomic reads. For the vast majority of heterozygous loci there was a linear correlation between the RNA-Seq signal and the inferred allele count. When only heterozygous sites corresponding to the allele ratio A:A:B:B were investigated, 59\% of the analyzed sites displayed expression imbalance. It can be assumed that most of the allelic imbalance is caused by allele dosage rather than \emph{cis} or \emph{trans} regulatory differences. In conclusion, the current data indicate that both nuclei are transcriptionally active, and there was a correlation between expression level and allele dosage. Together these data indicate that transcription in \emph{G. intestinalis} is symmetric.

\subsubsection{Only Six Genes are \emph{cis}-spliced}
Only five genes are currently reported to contain introns and undergo \emph{cis}-splicing in \emph{G. intestinalis}. These genes are listed in Table \ref{tab:introns}. We performed an exhaustive search for new \emph{cis}-splicing events using a comparative transcriptomics approach. Putative splice junctions of WB, AS175, P15 and GS were first mapped with the software TopHat \cite{Trapnell2009}. Manual inspection of the deduced splice junctions indicated a large number of dubious suggestions, as concluded from the proposed splice pattern and the annotation of the involved genes. For example, many multicopy genes were suggested to undergo extensive splicing, e.g. \emph{vsp} and \emph{p21.1} genes. The repetitive nature of these genes is likely to give artifacts when mapping short reads, especially since the implemented algorithms chop reads into even smaller pieces before mapping them. To increase the signal to noise ratio, the algorithmically identified splice sites were filtered according to these criteria: (\emph{i}) the splicing pattern was required to be conserved in at least two isolates; (\emph{ii}) repeated genes were discarded (e.g. \emph{vsp}, \emph{hcmp} and \emph{p21.1}); (\emph{iii}) the intron had to be confined to the ORF or the closest upstream intergenic region; and (\emph{iv}) the splice pattern had to be supported by a minimum of 5 reads.

\begin{table}[h]
\caption{Confirmed introns in \emph{G. intestinalis}}
\centering 
\begin{threeparttable}
\begin{tabular}{c rrrrrr} 
\hline\hline 
Gene & ID \tnote{a} & Ref. & Boundary \tnote{b} & nt \tnote{c} & \#Isolates \tnote{d} & Pos. \tnote{e} \\ [0.5ex]
\hline 
[2Fe-2S] ferredoxin & 27266 & \cite{Nixon2002} & CT-AG & 35 & 4 & 0.05 \\ 
\emph{Rpl7a} & 17244 & \cite{Russell2005,Morrison2007} & GT-AG & 109 & 4 & 0.54 \\ 
dynein light-chain & 15124 & \cite{Morrison2007} & GT-AG & 32 & 3 & 0.05 \\ 
uncharacterized & 35332 & \cite{Russell2005} & GT-AG & 220 & 2 & 0.01 \\ 
uncharacterized & 15604 & \cite{Roy2012} & GT-AG & 29 & 4 & 0.02 \\ 
uncharacterized & 86945 & novel & GT-AG & 36 & 4 & 0.02 \\
\hline 
\end{tabular}

\begin{tablenotes}
\footnotesize \item[a] Prefix: GL50803\_
  \item[b] The intron boundaries, 5$'$ to 3$'$ (splice sites).
  \item[c] Length of the intron in nucleotides.
  \item[d] Number of isolates the splicing pattern was found in.
  \item[e] Intron position in the gene along the 5$'$ to 3$'$ axis. The position was calculated as: [position of first nucleotide of the intron]/[gene length].
\end{tablenotes}

\label{tab:introns}
\end{threeparttable}
\end{table}

\FloatBarrier

The five previously reported introns were found in our data, indicating that the bioinformatics procedure was robust. Three of the previously confirmed splice variants were found in all 4 isolates, and two were found only in 3 and 2 isolates, respectively (Table \ref{tab:introns}). This reflects limitations in our data sets or genome assemblies rather than differential splicing between isolates. The bioinformatic search suggested 14 new intron candidates. However, PCR validation on genomic DNA and RNA (cDNA) subsequently rejected 13 of the 14 candidates, which suggests that splice site prediction using RNA-Seq is noisy. One new intron was confirmed by PCR, and the amplified DNA was sequenced with dye-terminator sequencing, confirming splice sites.

\begin{figure}[!ht]
 \centering
 \includegraphics[width=4.5in]{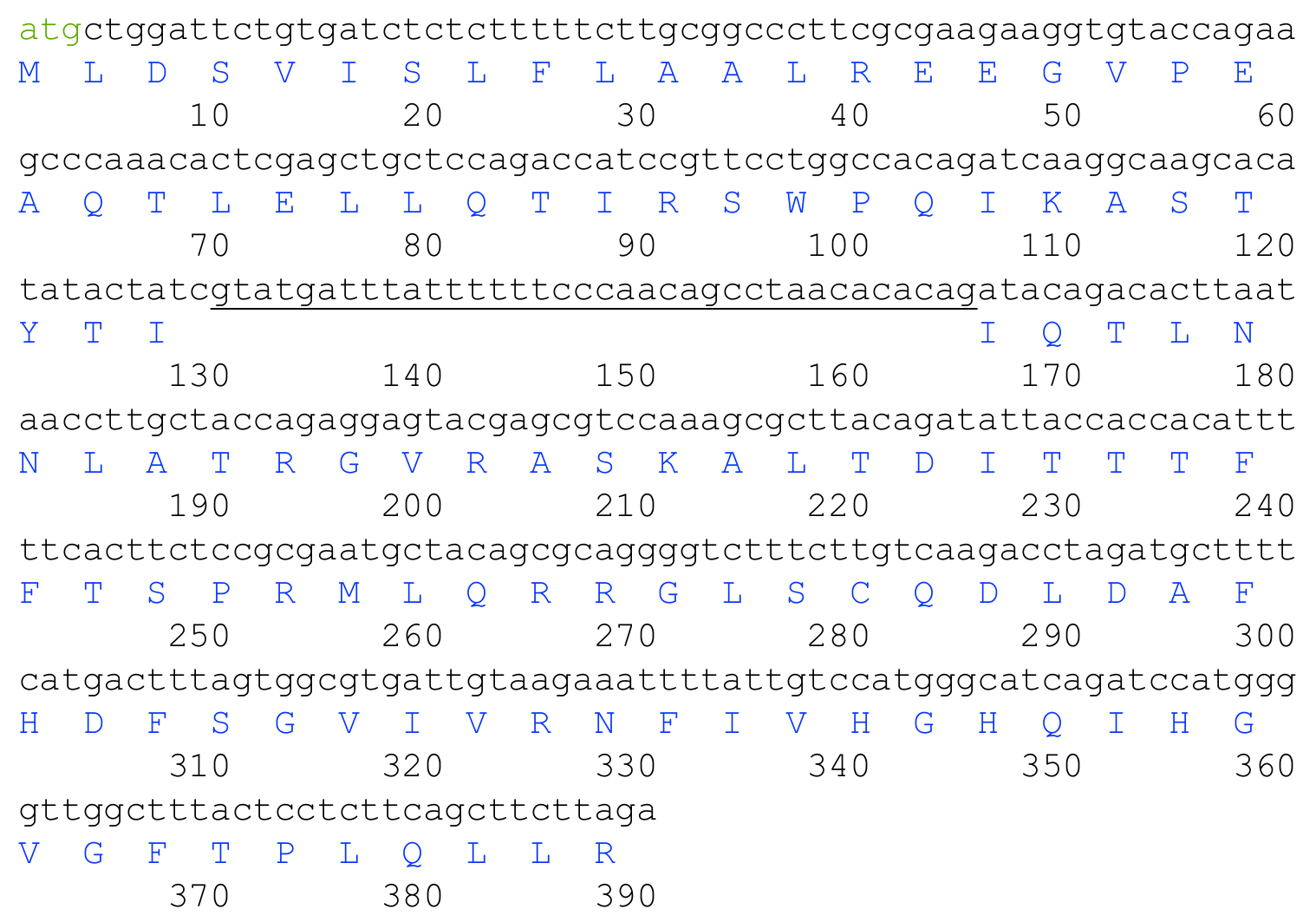}
  \caption{ \small 36-nt intron (underlined) in the gene GL50803\_86945, encoding an uncharacterized protein. Only the first 390 nt of the gene is shown. The translational start codon is indicated in green and the conceptual translation is shown in blue. }

\label{intron}
\end{figure}

The novel intron was 36 nt in length and was present in an uncharacterized gene on chromosome 4 (Table \ref{tab:introns}; Figure \ref{intron}). The intron was identified in all four isolates and had canonical splice boundaries (GT-AG). Removal of the intron extends the open reading frame with 73 codons. It is possible that some putative introns were missed, especially low-level \emph{cis}-splicing events. However, the true number of introns in \emph{G. intestinalis} is not likely to be much higher than presented here.

\FloatBarrier

Five out of six introns displayed a positional bias towards the 5' end of the gene (Table \ref{tab:introns}). Intron deficit and 5$'$ bias have been observed in the two protozoa \emph{Encephalitozoon cuniculi} \cite{Katinka2001} and \emph{Guillardia theta} \cite{Douglas2001}, both of which belong to intron-rich groups of organisms. Gene sampling of species of the order \emph{Oxymonadida} (anaerobic flagellates found mainly in the gut of termites and wood-eating roaches) have found evidence of protozoa with extensive intron content \cite{Slamovits2006}. These data corroborate the idea that \emph{G. intestinalis} derived from a more intron-rich ancestor but lost introns during evolution.

\begin{center}
\decofourleft\decofourright
\end{center}

\newpage 


\section{Concluding remarks}
\fancyhead[RO,LE]{}

\setlength\epigraphwidth{3.5in}
\setlength{\epigraphrule}{0pt} 

\epigraph{\emph{There is grandeur in this view of life, with its several powers, having been originally breathed into a few forms or into one; and that, whilst this planet has gone cycling on according to the fixed law of gravity, from so simple a beginning endless forms most beautiful and most wonderful have been, and are being, evolved.}}{-- C. Darwin (1809-1882), \emph{On the Origin of Species}.}

\vspace{0.6cm}

\subsubsection{Comparative Genomics}
The sequence data generated in the present studies are publicly available in integrative databases together with data sets from independent studies. Much of the data have already been utilized in several independent studies (for example \cite{Manning2011,Williams2011,Chen2011,Feng2012,Xu2012}), which underscore the value of large genomic data sets in parasitology: both to increase understanding of parasite biology and for hypothesis generation. As sequencing technologies continue to improve in terms of output and costs, it will be of value for the research community to undertake large-scale efforts to sequence multiple genomes of biologically relevant strains. Similar to the 1000 genomes project \cite{Clarke2012}, future efforts in parasitology may target several hundred or more strains. The integration of such data sets into community-oriented databases, for example EuPathDB \cite{Aurrecoechea2010}, should allow researchers to take advantage of the vast amount of data. One challenge for the future will be to develop suitable phenotyping strategies for protozoan parasites, since sequence data from many strains may be of limited value if phenotypes are unknown. Second, we can learn about the evolutionary adaptations that led to parasitism from genome comparisons of even more diverged species, for example free-living or avirulent species within the diplomonadida and kinetoplastida. Examples of such species are \emph{Spironucleus sp.} (diplomonadida) and \emph{Trypanosoma rangeli} (kinetoplastida). Such sequencing projects are currently being undertaken and are likely to yield new insights into parasite evolution.
\newline \indent Many genes of these parasites are currently uncharacterized. A future priority of parasitology  should therefore be to explore the functionality of uncharacterized genes, since these are the genes likely to be parasite-specific rather than universally conserved among eukaryotes. In theory, it would also be possible to find drug targets among these genes. In conclusion, many questions relating to the basic biology and evolutionary history of these parasites are still unresolved. For example, do isolate-specific genes contribute to the phenotype? What is the mechanism behind horizontal gene transfer and how often does it happen? Why do some parasite lineages contain higher heterozygosity than others? Do protozoan parasites recombine and how often? Genome sequencing may not by itself provide answers to these questions; rather genomic data should be used together with functional studies or other data sets. Ultimately, this may facilitate new insights into the basic biology that may lead to new treatments and control of these pathogens.

\subsubsection{The Short Transcriptome of \emph{T. cruzi}}
There was no evidence of canonical small RNAs as often found in metazoans, e.g. microRNAs, small interfering RNAs or piwi-interacting RNA. This finding is consistent with the lack of RNA interference in \emph{T. cruzi}. More than 90\% of small RNAs were derived from known non-coding RNAs (tRNA, rRNA and snRNA). The most prominent category was tRNA-derived small RNAs. It remains to be determined if these small RNA classes are functional, partially functional or merely debris from the RNA ``degradome,'' i.e. turnover. The following observations warrant further investigation of \emph{T. cruzi} small RNAs: (\emph{i}) the cleavage pattern appears to be non-random; (\emph{ii}) certain small RNAs appear to be stable, as suggested by the copy number; (\emph{iii}) tsRNA locates to cytoplasmic granules \cite{GarciaSilva2010}; and (\emph{iv}) certain tRNA isoacceptors were overrepresented in terms of deriving tsRNAs. Interestingly, we identified a population of small RNAs that were not derived from known non-coding RNAs, and were predicted to contain microRNA-like seed regions. The present study only briefly scratches the surface of the small RNA transcriptome and raises questions that call for further investigation.

\subsubsection{Comparative Transcriptomics of \emph{G. intestinalis}}
Transcription is one key event in the translation of genotype to phenotype, and transcriptome studies can enhance our understanding of phenotypic differences and the evolutionary trajectories of pathogens. Almost the entire genome of \emph{G. intestinalis} was transcribed in trophozoites grown \emph{in vitro}, confirming the promiscuous nature of transcription in this parasite. The data confirmed many gene models that were originally annotated without transcriptional evidence, and also identified novel protein-coding genes. The extent of gene expression divergence recapitulated the known phylogeny of the strains, suggesting that gene expression has largely evolved via genetic drift. Despite many genes with strain-specific expression, it is difficult or impossible to conclude what differences are of functional importance. Furthermore, it is important to note that differences in mRNA abundance may not necessarily result in differences at the protein level.

Notably, a non-expressed gene locus was identified, consisting of 28 non-transcribed genes, which may reflect transcriptional repression by a yet-to-be described mechanism. Sequence signatures indicated a putative viral origin. Perhaps silencing operates at the level of chromatin organization. Functional investigation will be required to elucidate the mechanism behind this mode of silencing.

Biallelic transcription was identified at most of the heterozygous loci of the \emph{G. intestinalis} isolate GS, which were identified and described in paper \emph{i}. Comparison of allele dosage with allelic transcription levels indicated a relationship between the two variables. The data corroborate previous observations of binucleic transcription in \emph{G. intestinalis} and further provide an association between allele dosage and transcription levels. These data suggest that binucleic transcription is largely symmetric, which nonetheless does not preclude the existence of allele-specific expression. However, the latter does not appear to be a general feature of transcription in \emph{G. intestinalis}, which is also consistent with the deficit of regulation at the transcriptional level.

Previously reported introns were confirmed and only one new intron was discovered, suggesting that the true number of introns is not likely to be much higher than this. It is tempting to speculate that \emph{G. intestinalis} may have been more intron-rich in the past, but undergone intron-loss. Conversely, it is also possible that introns became more prevalent in eukaryotes after the split of \emph{G. intestinalis} from the main eukaryotic lineage. However, it seems unlikely that six introns have necessitated the evolution of the relatively complex splicing machinery of \emph{G. intestinalis}. The final evidence to settle this question would be the finding of a diplomonad relative with an extensive repertoire of introns, which is yet to be discovered.

\begin{center}
\decofourleft\decofourright
\end{center}

\newpage 

\section{Populärvetenskaplig sammanfattning}
\fancyhead[RO,LE]{}
\fancyhead[LO,RE]{}

De två parasiterna \emph{Trypanosoma cruzi} och \emph{Giardia intestinalis} är encelliga organismer som orsakar sjukdom och lidande hos flera miljoner människor världen över. \emph{Trypanosoma cruzi} är huvudsakligen ett problem i Latinamerika där den ger upphov till Chagas sjukdom. {\raise.17ex\hbox{$\scriptstyle\sim$}}8 miljoner människor är infekterade i Latinamerika och 11,000 avlider varje år till följd av sjukdomen \cite{Rassi2010}. I Sverige finns grovt uppskattat 1,000 infekterade personer \cite{Sandahl2011}. \emph{Giardia} drabbar människor runt om i hela världen och kan leda till diarrésjukdomen giardiasis. Migration leder till att sjukdomarna blir vanligare i Europa, Nordamerika och andra delar av världen. Både Chagas sjukdom och \emph{Giardia} räknas som försummade sjukdomar, och orsakar problem i utvecklingsländer \cite{Savioli2006,Rassi2010}. De nämnda sjukdomarna, tillsammans med flera andra tropiska sjukdomar, prioriteras ofta inte i fråga om läkemedelsutveckling eftersom det inte anses ekonomiskt lönsamt.

I den här avhandlingen har jag använt datoranalyser för att studera biologisk information från de två nämnda parasiterna, bland annat gener som kodas i dess arvsmassa. Information från parasiterna avläses med särskilda instrument som med hög noggrannhet avläser den genetiska koden. Därefter analyseras informationen för att hitta biologisk relevanta mönster som kan öka förståelsen av parasiternas biologi. Detaljerade analyser och jämförelse mellan olika stammar beskrivs, och identifierar egenskaper som är kodade i parasiternas DNA. De metoder som använts avslöjar också evolutionära mönster, som kan bidra till att öka förståelsen för hur parasiterna anpassat sig till människan och hur de undviker immunförsvaret. I den här avhandlingen har även genernas aktivitet studerats, det vill säga genernas uttryck. Genuttrycket kan avslöja hur olika stammar skiljer sig funktionellt. Information från de genomförda studierna kan i framtiden användas för att designa mer skräddarsydda experiment som kan leda till bättre behandlingsmetoder och diagnostik.

\newpage 

\section{Acknowledgments}

\vspace{0.3cm}

\setlength\epigraphwidth{3in}
\setlength{\epigraphrule}{0pt} 
\epigraph{\emph{Skepticism is the agent of reason against organized irrationalism -- and is therefore one of the keys to human social and civic decency.\newline}}{-- S.J. Gould (1941-2002), evolutionary biologist.}

\vspace{1.0cm}

\noindent I'm grateful to have been given the opportunity to do a PhD and for the freedom to be creative. I want to thank and acknowledge friends and colleagues I met during my PhD studies, without whom the journey would have been less fun and fulfilling.\newline

\noindent My supervisor prof. \textbf{Björn Andersson} -- for scientific and moral support and for also allowing me to learn science abroad. My co-supervisor prof. \textbf{Staffan Svärd} and his group – for good science, support and travel company.\newline

\noindent In addition, I acknowledge the following people (the names are not in any particular order): \textbf{Stephen Ochaya} -- for chats about science and life. \textbf{Jon Jerlström-Hultqvist} -- for experiments, discussions and collaboration, and for travel company to Kampala and Wellington among others. \textbf{Marcela Ferella} -- for experiments, discussions and collaboration. \textbf{Lena Åslund} -- for discussions on \emph{T. cruzi} small RNAs. \textbf{Erik Arner} -- for sharp bioinformatics skills, \textbf{Carsten Daub} -- for allowing me to visit RIKEN. \textbf{Morana}. \textbf{Ellen Sherwood} -- for answering questions during my first year. \textbf{Öjar Melefors}, \textbf{Alma Brolund}, \textbf{Linus Sandegren} and \textbf{Karin Tegmark Wisell} for interesting discussions about plasmid research. \textbf{Karin Troell} and \textbf{Johan Ankarklev}, for scientific discussions. \textbf{Rickard Sandberg} -- for doing great research and allowing me to use the cluster. \textbf{Elin Einarsson} for 3$'$ RACE. \textbf{Bengt Persson} for runtime at LiU. \textbf{Elsie Castro} for running 454. \textbf{Nicolas} and \textbf{Emilie} for friendship. \textbf{Carlos Talavera-López} for PCRs. My co-authors at LSHTM and UEA: Prof. \textbf{Michael Miles}, \textbf{Claire Butler}, \textbf{Louisa Messenger}, \textbf{Michael Lewis}, \textbf{Martin Llewellyn} and \textbf{Kevin Tyler}. \textbf{Anna Wetterbom} for chats and advice. \textbf{Hamid Darban}, \textbf{Stefanie Prast-Nielsen}, \textbf{Johannes Luthman} for chats about science and life. \textbf{Jan Andersson} and \textbf{Feifei Xu} for evolutionary insights. \textbf{Fredrik Lysholm} for technical chats. \textbf{Daniel Nilsson} for answering questions,  especially during my first year. The fellow PhD students I met on the comparative genomics course 2011 at Pasteur Institute. And to anyone I forgot: I apologise, it is entirely my fault.\newline

\noindent I also acknowledge sources of funding; \textbf{KID-medel}, \textbf{FORMAS}, \textbf{Vetenskapsrådet}, as well as the people that were not directly involved in this piece of research: administrators, technical staff, annotators, anonymous reviewers, open source developers and all others that work to create a good research environment.\newline

\noindent I acknowledge friends and family outside of research. \textbf{Robert} and \textbf{Christoffer} for long-time friendship. \textbf{Magnus}, \textbf{Henrik} and their spouses for always inviting me. My sister and my parents: \textbf{Fanny}, \textbf{Peringe} and \textbf{Sinikka} for support.\newline

\noindent Now that I reached the last lines I would like to wish past, present and future friends all the best of luck.\newline

\begin{center}

Oscar, \today. Stockholm.\\[0.7cm]

\includegraphics[width=0.2\textwidth]{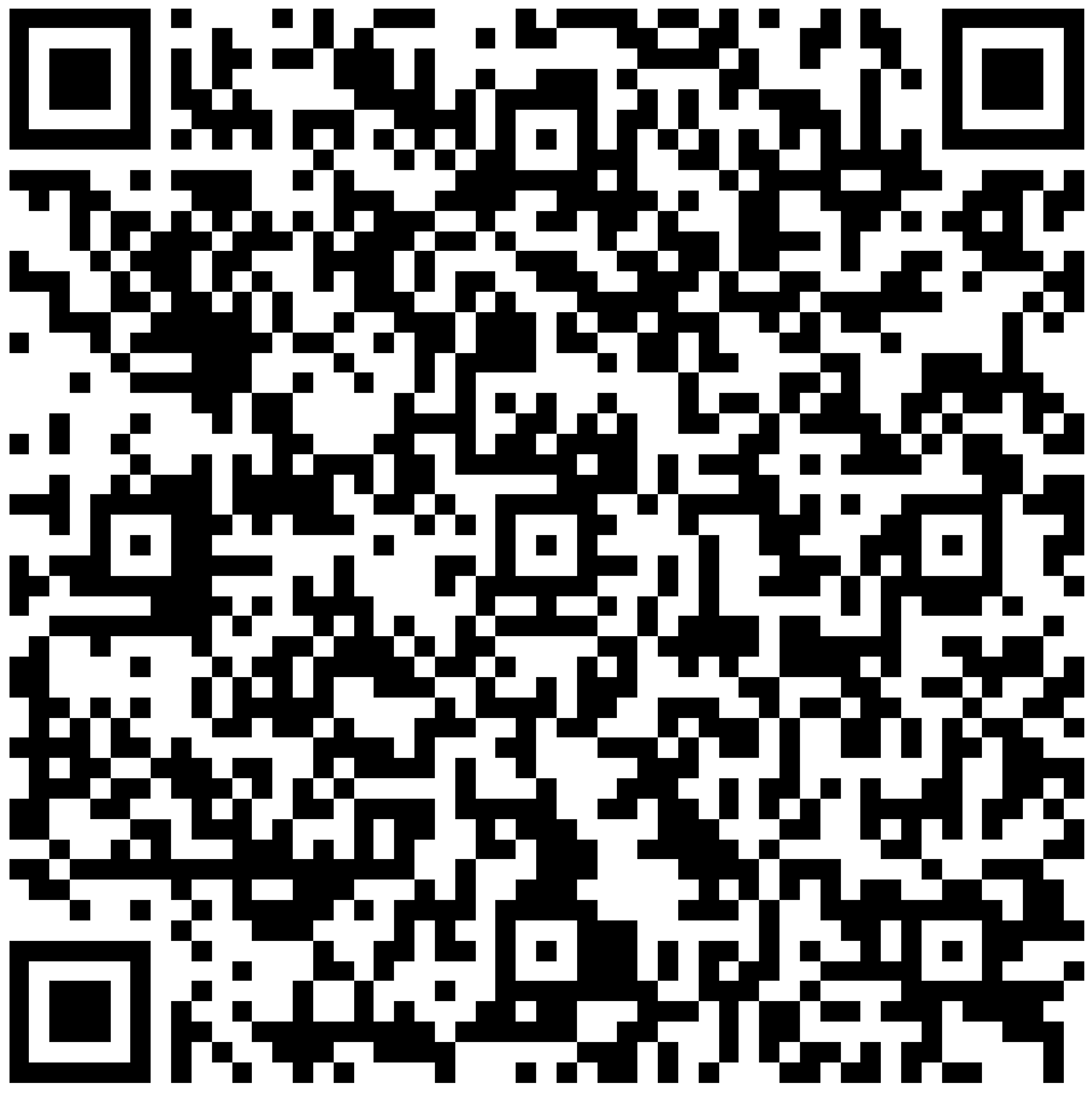}\\[0.5cm]

\small This thesis was typeset with \LaTeX{}.
\end{center}

\begin{center}
\decofourleft\decofourright
\end{center}


\newpage 

\nocite{*}
\section{References}

\setlength\epigraphwidth{3in}
\setlength{\epigraphrule}{0pt} 
\epigraph{\emph{Nothing in life is to be feared, it is only to be understood. Now is the time to understand more, so that we may fear less.\newline}}{-- M.S. Curie (1867-1934), physicist and chemist.}

\vspace{1.0cm}

\renewcommand*{\bibfont}{\footnotesize}
\printbibliography[heading=none,type=article]

\begin{center}
\decofourleft\decofourright
\end{center}

\end{document}